\documentclass[aps,prd,onecolumn,nofootinbib,superscriptaddress,groupedaddress]{revtex4}  
\usepackage{graphicx}  
\usepackage{dcolumn}   
\usepackage{bm}        
\usepackage{amssymb}   
\usepackage{threeparttable} 
\usepackage{subfig}  
\usepackage{changepage}
\usepackage{amsmath}
\usepackage{color}
\usepackage{ifthen}
\usepackage[pdftex]{hyperref}
\usepackage{leftidx}

\newcommand{\lensEstbiased}{\hat{C}^{dd(\text{biased})}}
\newcommand{\lensEstcutbiased}{\hat{C}^{dd(\text{cut})}}

\newcommand{\CEobs}[1]{C^{EE}_{{\rm obs}, #1}}
\newcommand{\CBobs}[1]{C^{BB}_{{\rm obs}, #1}}
\newcommand{\Cobs}[1]{C_{{\rm obs}, #1}}

\newcommand{\CEnoise}{N^{EE}}
\newcommand{\CBnoise}{N^{BB}}
\newcommand{\Cnoise}{N}

\newcommand{\Cdinput}{C^{dd}}
\newcommand{\NZero}{N^{(0)}}
\newcommand{\NOne}{N^{(1)}}
\newcommand{\philens}{\phi}
\newcommand{\Cphiphi}{C^{\philens\philens}}

\newcommand{\hd}{\hat{d}}

\newcommand{\Nsim}{n_{\rm sim}}
\newcommand{\pseudoE}{E^{\rm pseudo}}
\newcommand{\pseudoB}{B^{\rm pseudo}}
\newcommand{\fsky}{f_{\rm sky}}

\newcommand{\muK}{\mu\rm{K}}
\newcommand{\arcmin}{{\rm arcmin}}
\newcommand{\muKarcmin}{\,\muK\arcmin}

\providecommand{\CAMB}{\textsc{camb}}

\newcommand{\begm}{\begin{pmatrix}}
\newcommand{\enm}{\end{pmatrix}}

\newcommand\ba{\begin{eqnarray}}
\newcommand\ea{\end{eqnarray}}
\newcommand\bea{\begin{eqnarray}}
\newcommand\eea{\end{eqnarray}}

\newcommand\be{\begin{equation}}
\newcommand\ee{\end{equation}}

\newcommand{\vgrad}{{\boldsymbol{\nabla}}}

\newcommand{\la}{\langle}
\newcommand{\ra}{\rangle}

\newcommand{\boldvec}[1]{{{\mathbf{#1}}}}

\newcommand{\vL}{\boldvec{L}}

\newcommand{\vd}{\boldvec{d}}

\newcommand{\vl}{\boldvec{l}}

\newcommand{\vn}{\boldvec{n}}

\newcommand{\vx}{\boldvec{x}}

\newcommand{\clo}{\mathcal{O}}

\newcommand{\vnhat}{\hat{\vn}}

\def\eprinttmp@#1arXiv:#2 [#3]#4@{
\ifthenelse{\equal{#3}{x}}{\href{http://arxiv.org/abs/#1}{#1}}{\href{http://arxiv.org/abs/#2}{arXiv:#2} [#3]}}

\providecommand{\eprint}[1]{\eprinttmp@#1arXiv: [x]@}
\newcommand{\adsurl}[1]{\href{#1}{ADS}}
\providecommand{\bibinfo}[2]{\ifthenelse{\equal{#1}{isbn}}{
\href{http://cosmologist.info/ISBN/#2}{#2}}{#2}}

\begin{document}

\title{CMB lensing reconstruction using cut sky polarization maps and pure-$B$ modes}
\date{\today}

\author{Ruth Pearson}
\address{Department of Physics \& Astronomy, University of Sussex, Brighton BN1 9QH, UK}
\address{Kavli Institute for Particle Astrophysics and Cosmology, SLAC, 2575 Sand Hill Road, Menlo Park, CA 94025, USA}

\author{Blake Sherwin}
\address{Department of Physics, University of California, Berkeley, CA 94720, USA}
\address{Miller Institute for Basic Research in Science, University of California, Berkeley, CA 94720, USA}

\author{Antony Lewis}
\address{Department of Physics \& Astronomy, University of Sussex, Brighton BN1 9QH, UK}

\begin{abstract}
 Detailed measurements of the CMB lensing signal are an important scientific goal of ongoing ground-based CMB polarization experiments, which are mapping the CMB at high resolution over small patches of the sky. In this work we simulate CMB polarization lensing reconstruction for the $EE$ and $EB$ quadratic estimators with current-generation noise levels and resolution, and show that without boundary effects the known and expected zeroth and first order $\NZero$ and $\NOne$ biases provide an adequate model for non-signal contributions to the lensing power spectrum estimators.
Small sky areas present a number of additional challenges for polarization lensing reconstruction, including leakage of $E$ modes into $B$ modes. We show how simple windowed estimators using filtered pure-$B$ modes can greatly reduce the mask-induced mean-field lensing signal and reduce variance in the estimators. This provides a simple method (used with recent observations) that gives an alternative to more optimal but expensive inverse-variance filtering.

\end{abstract}

\maketitle

\section{Introduction}
Photons of the Cosmic Microwave Background (CMB) are deflected by gravitational potentials along the line of sight as they travel from the surface of last scattering to our telescopes. This gravitational lensing by large scale structure leads to a remapping of the observed CMB sky by $\approx 3$ arcminutes (RMS). To achieve the most precise measurements of the primary CMB (such as measurements of $B$ mode polarization induced by primordial tensor fluctuations), these CMB lensing deflections must be modeled accurately and corrected for~\cite{Hu:2001fb,Lewis:2001hp}.  However, the CMB lensing signal is also of great interest for cosmology in itself, because it traces structure from the surface of last scattering until today and is thus a powerful probe of the matter distribution (and hence also of the properties of neutrinos and dark energy). While the surface brightness of the CMB is preserved by the lensing remapping of individual photons on the sky, this remapping alters the statistical properties of the observed CMB anisotropies.  For example, CMB lensing induces lensing $B$ modes as well as non-Gaussianity (when averaged over realizations of large scale structure) in the data (for reviews see Refs.~\cite{Lewis:2006fu,Hanson:2009kr}). For the one particular distribution of matter in our universe, the statistical effect of lensing appears instead as off-diagonal covariances between modes in the CMB observables. These lensing-induced covariances can be used to reconstruct the lensing potential~\cite{Hu:2001kj}.

 To reconstruct the lensing potential, a full maximum likelihood based analysis is most optimal~\cite{Hirata:2003ka}. However this is computationally challenging, and a leading-order quadratic estimator approximation is usually used instead \cite{Hu:2001kj,Okamoto03}. These estimators are nearly optimal for current-generation experiments once generalized for partial sky coverage and inhomogeneous noise~\cite{Hirata:2002jy,Hirata:2003ka,Hanson:2009gu,Hanson:2009kr}.
 On the cut sky slightly less optimal but simpler estimators can also be used which use apodized sky maps without the numerically-expensive full inverse-variance weighting required for the perturbatively optimal estimators. We focus on these simpler estimators, as used by various current-generation ground-based experiments, as described in Section~\ref{sec:lensintro}.

The temperature ($TT$) quadratic estimator (consisting of a quadratic combination of two temperature modes) has been used to measure the CMB lensing potential to high significance~\cite{Das:2011ak, Das:2013zf, vanEngelen:2012va}, most recently at more than $20\sigma$ by the Planck collaboration~\cite{Ade:2013tyw}.  The signal-to-noise ratio for lensing reconstruction from CMB polarization data is expected to be much better in the future, because polarization lensing is not limited by cosmic variance, with $B$ modes on small scales expected to be vanishingly small on the unlensed sky.  First examples of CMB lensing reconstruction from polarization data use SPT or POLARBEAR data in cross- or auto-correlation~\cite{Hanson:2013hsb,Ade:2013gez,Ade:2013hjl}, based on CMB polarization observations on small patches of the sky at high resolution.

In this work we investigate the ability of such current and next-generation polarization observations to measure the CMB lensing potential power spectrum.  We begin by studying periodic boundary conditions for the $EE$ and $EB$ quadratic estimators in section~\ref{sec:fullsky}.  We show that the known $\NZero$ and $\NOne$ power spectrum biases are sufficient to model the reconstructed lensing potential power spectrum.
Since no experiment can actually measure the full sky, we also consider the effect of using cut sky maps from small patches of sky.
Because the mapping between the observed polarization Stokes parameters ($Q$ and $U$) and the physical $E$ and $B$ polarization fields is non-local (involving derivatives), on a patch of the sky $E$ modes can leak into $B$ modes. This provides an additional complication for lensing reconstruction, and could, if not mitigated, significantly impair the use of $B$ modes to reconstruct the lensing field.
 In section~\ref{sec:pureBintro} we outline the effect and review the `pure' $B$ mode construction that can be used to project the observed data into clean $B$ modes that we then incorporate into our lensing reconstruction pipeline.  In section~\ref{sec:cutsky} we test the $EE$ and $EB$ quadratic estimators when applied to $4.5^{\circ}\times4.5^{\circ}$ patches of sky with non-periodic boundaries, and assess whether using pure-$B$ modes can help to improve the reconstruction errors. We also assess the magnitude of any additional biases in the cut-sky case.

 \subsection{CMB Lensing}
 \label{sec:lensintro}
 Since we are mainly interested in small patches of sky for this paper, we will use the flat-sky approximation, following the notation of \cite{Hu:2001kj}.  The temperature $T$ at a position $\vx$ on the plane of the sky is then expanded into harmonics as
 \be
T(\vx)=\int\frac{d^{2}\vl}{(2\pi)^{2}}T(\vl) e^{i\vl \cdot \vx}.
 \ee
The observed Q and $U$ Stokes parameters are expressed in terms of $E$ and $B$ polarization modes as

 \begin{equation}
\label{ }
[Q\pm iU](\vx)=-\int\frac{d^{2}\vl}{(2\pi)^{2}}[E(\vl )\pm iB(\vl )] e^{\pm2i\varphi_{l}}e^{i\vl \cdot \vx},
\end{equation}
where the plane wave vector $\vl $ is the flat-sky analogue of the full-sky spherical harmonic $lm$, and $\cos \varphi_{l}=\hat{\textbf{x}} \cdot \hat{\vl }$.  The fields at last scattering are re-mapped by the lensing as
\begin{equation}
\label{ }
T(\vx)=\tilde{T}(\vx +\vd(\vx)), \qquad
[Q\pm iU](\vx)=[\tilde{Q}\pm i\tilde{U}](\vx +\vd(\vx)),
\end{equation}
where the tilde denotes unlensed fields and $\vd(\vx)$ is the lensing deflection field.  The deflection can be expressed as $\vd =\vgrad \philens$, where in a flat FRW universe the lensing potential $\philens$ is given by
\begin{equation}
\label{ }
\philens(\vx)=-2\int_0^{\chi_*} d\chi \frac{(\chi_{*}-\chi)}{\chi\chi_{*}}\Psi(\chi\vnhat_\vx).
\end{equation}
Here $\chi$ is the comoving distance along the line of sight, $\chi_{*}$ is the distance to the last scattering surface, and
$\Psi(\chi\vnhat_\vx)$ is the (Weyl) gravitational potential at the photon location along the line of sight in direction $\vnhat_\vx$ on the sky.

Lensing of the CMB can be measured from the response of the lensed two point correlation function to the lensing potential.
We have multiple fields, so in general have multiple quadratic combinations to consider, $X_i(\vl_1) X_j(\vl_2)$, where $X_i$ could be $T$, $E$ or $B$. Considering the lensing potential $\philens$ to be fixed, averaging over all other modes and neglecting correlations between the lensing potential and CMB, to linear order in $\phi$
\be
\la  X_i(\vl_1) X_j(\vl_2) \ra_{\philens} \approx
\int d\vL' \left\la  \frac{\delta }{\delta \philens(\vL')}\left(X_i(\vl_1) X_j(\vl_2) \right)\right\ra \philens(\vL') =
 f_{ij}(\vl_1,\vl_2) \philens(\vL),
\ee
where $\vl_1+\vl_2=\vL$ and $\vL\ne 0$. Here the response functions $f_{ij}$ are defined as the average linear response to a lensing mode $\phi(\vL)$~\cite{Lewis:2011au,Lewis:2011fk}
\be
\left\la \frac{\delta }{\delta \philens(\vL)}\left(X_i(\vl_1) X_j(\vl_2) \right)\right\ra=  \delta(\vl_1+\vl_2-\vL)f_{ij}(\vl_1,\vl_2).
\ee
For a result to leading order in the particular mode $\philens(\vL)$, the expectation can be evaluated over all the fields (the unlensed CMB, and non-zero lensing modes that are also present); the result is then proportional to a delta-function by statistical homogeneity.
To zeroth order in the lensing potential the response functions $f_{ij}$ are given by Ref.~\cite{Hu:2001kj}, however because the lensing is a substantial signal, to get the normalization right higher order corrections must be included~\cite{Hanson:2010rp}, corresponding to including the contribution of other lensing modes to the covariance\footnote{An $\clo(\Cphiphi)$ correction to the power spectrum normalization, giving a total error $\clo((\Cphiphi)^2)$, and hence an $N^{(2)}$ if neglected~\cite{Hanson:2010rp}.}. Explicit exact expressions (for Gaussian unlensed fields) are given in~\cite{Lewis:2011fk}, and can be approximated quite accurately (non-perturbatively) by using the lensed CMB power spectra in place of the unlensed ones in the results of~\cite{Hu:2001kj}: to good approximation when we look for a mode $\philens(\vL)$, the change induced on the sky depends on how it affects the lensed CMB, where the lensing effect is dominated by lensing from other modes that are also present. In the case of polarization the main non-perturbative effect that should be modelled is the substantial effect of lensing on $EE$. There are also additional corrections of $\clo(C_l^{BB})$, but these are much smaller (just including the lensed $C_l^{BB}$ does not include all the terms of equivalent order~\cite{Lewis:2011fk}). For current observations corrections of $\clo(C_l^{BB}/C_l^{EE})$ can probably be safely neglected; further perturbative tests of the lensed-$C_l$ approximation are given in Ref.~\cite{Anderes:2013jw}.

Weighting functions $W(\vl_1,\vl_2)$ can be used to sum the quadratic combinations $X_i(\vl_1) X_j(\vl_2)$ so that the deflection field estimators are
\be
\hd_{ij}(\vL) = \frac{A_{ij}(L)}{L} \int\frac{d^{2}\vl_1}{(2\pi)^{2}}X_i(\vl_1)X_j(\vl_2)W_{ij}(\vl_1,\vl_2),
\ee
where $\vl_1+\vl_2=\vL$, and $A_{ij}(L)$ is a normalization that makes the estimator unbiased when averaged over other modes:
\be
A_{ij}(L)=L^{2} \left[\int \frac{d^2\vl_1}{(2\pi)^{2}} f_{ij}(\vl_1,\vl_2) W_{ij}(\vl_1,\vl_2) \right]^{-1}.
\ee

 We can construct a naive estimator for the lensing power spectrum by measuring the power spectrum of the lensing deflection estimator. On the full sky this has an expectation value equal to the lensing potential power spectrum added to `noise' bias terms:
\be
\label{fullpow}
\la \hd^{*}_{ij}(\vL)\hd_{pq}(\vL ') \ra =(2\pi)^{2}\delta(\vL-\vL ')\left[C_{L}^{dd}+\NZero_{ijpq}(L) + \NOne_{ijpq}(L)\right]
\ee
to linear order in $C_L^{dd}$. Using the $\hd_{EE}(\vL)$ and $\hd_{EB}(\vL)$ quadratic estimators, there are 3 different ways to reconstruct the lensing power spectrum:  $\la \hd^{*}_{EE}(\vL)\hd_{EE}(\vL ') \ra$, $\la \hd^{*}_{EB}(\vL)\hd_{EB}(\vL ') \ra$ and $\la \hd^{*}_{EB}(\vL)\hd_{EE}(\vL ') \ra$.  The Gaussian $\NZero_{ijpq}(L)$ disconnected term is given by~\cite{Hu:2001kj}
\be
\NZero_{ijpq}(L) = \frac{A^{ij}(L)A^{pq}(L)}{L^2}\int \frac{d^2\vl_1 }{(2\pi)^2} W_{ij}(\vl_1,\vl_2)\left[
W_{pq}(\vl_1,\vl_2)\Cobs{l_1}^{ip}\Cobs{l_2}^{jq} +  W_{pq}(\vl_2,\vl_1)\Cobs{l_1}^{iq}\Cobs{l_2}^{jp}\right],
\ee
where $\Cobs{l}^{ij}$ are the total observed lensed CMB power spectra including (isotropic) instrumental noise.
In the diagonal case this simplifies to $\NZero_{ijij}(L) = A_{ij}(L)$ for optimized weights.  For the  $\la \hd^{*}_{EB}(\vL)\hd_{EE}(\vL ') \ra$ case $\NZero_{EBEE}=0$.

The $\NZero$ bias corresponds to random fluctuations in the noise and un-deflected CMB happening to look like lensing, and has contributions from both the Gaussian lensed power spectrum and instrumental noise.
The $\NOne_{ijpq}(L)$ term~\cite{Kesden:2003cc} is an additional variance from first order effects in the lensing power spectrum, given by
\be
 \NOne_{ijpq}(L) = \frac{A_{ij}(L)A_{pq}(L)}{L^2}\int \frac{d^2\vl_1 d^2\vL'}{(2\pi)^4} \Cphiphi_{L'}
 W_{ij}(\vl_1,\vl_2)\left[ f_{ip}(\vl_1,\vl_3) f_{jq}(\vl_2,\vl_4) W_{pq}(\vl_3,\vl_4) +
        f_{iq}(\vl_1,\vl_3) f_{jp}(\vl_2,\vl_4) W_{pq}(\vl_4,\vl_3) \right]
        \label{Nonedef}
\ee
where $\vl_1+\vl_2=\vL=-(\vl_3+\vl_4)$, $\vl_1+\vl_3=\vL' = -(\vl_2+\vl_4)$.
These expressions show the biases in the power spectrum of the deflection field $C_{l}^{dd}$, which is related to the power spectrum of the lensing potential $C_{l}^{\phi\phi}$ and power spectrum of the lensing convergence $C_{l}^{\kappa\kappa}$ by $C_{l}^{\kappa\kappa}=\frac{l(l+1)}{4}C_{l}^{dd}=\frac{l^{2}(l+1)^{2}}{4}C_{l}^{\phi\phi}$ on the full sky, and similarly with $l(l+1)\rightarrow l^2$ in the flat sky approximation.

\begin{figure}
\includegraphics[width=5in]{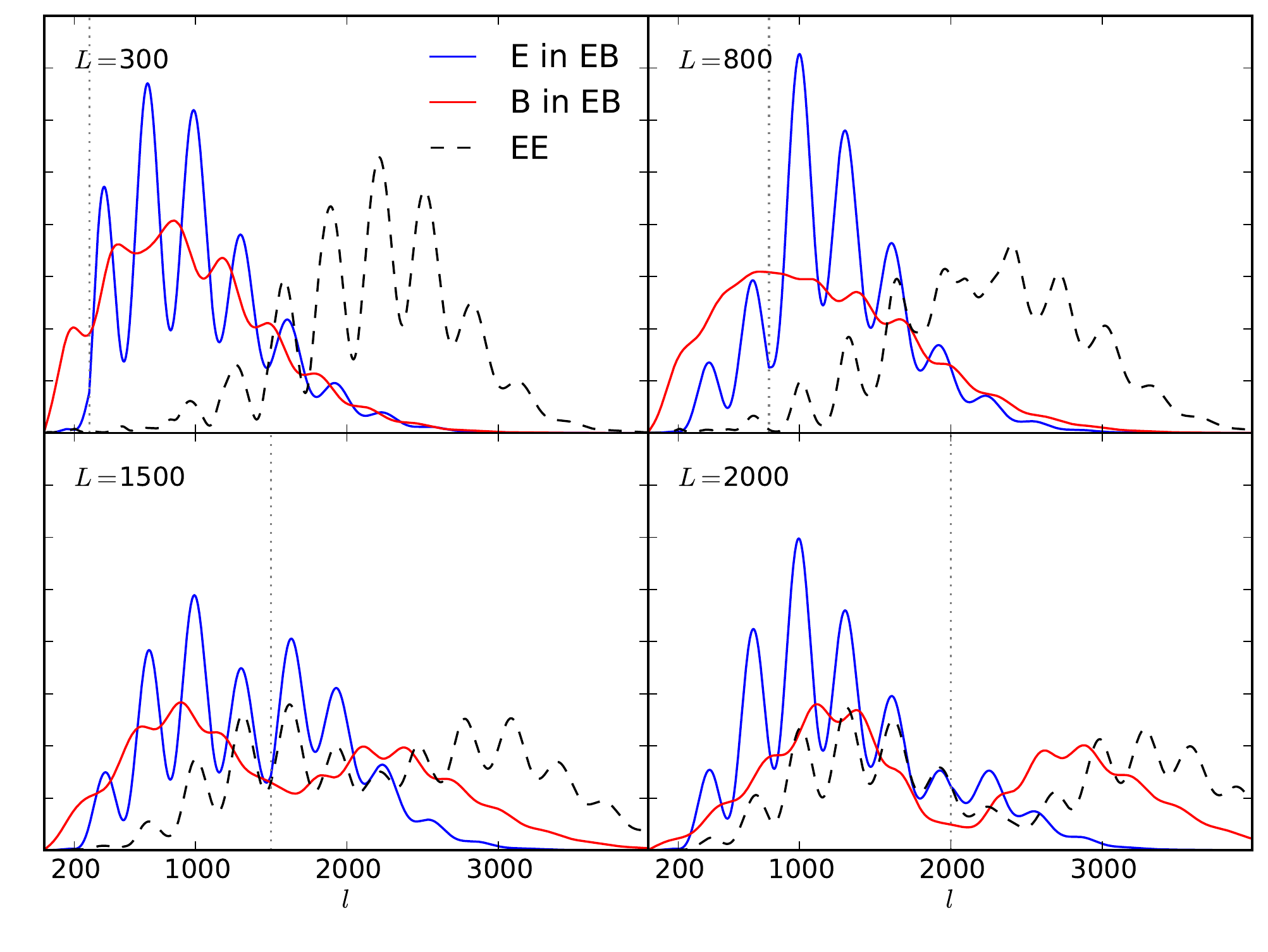}
\caption{Fractional contributions from $E(\vl)$ and $B(\vl)$ at $l=|\vl|$ to the lensing reconstruction at $L \in \{300, 800, 1500, 2000\}$ (four panels, where in each panel $l=L$ is marked with a dotted line), for the fiducial noise and resolution used in this paper. At the lower $L$ the $EE$ reconstruction (dashed lines) is mainly from squeezed shapes with $l\gg L$, however the $EB$ estimator the $E-$ and especially $B$-mode signal is important at much lower $l$ (solid lines). Mathematically what is plotted is $A_{ij}(L,l_1)\propto \int l_1 d\varphi_{l_1} f_{ij}(\vl_1, \vl_2) W_{ij}(\vl_1,\vl_2)$ as a function of $l_1$, or equivalently for $l_2$ in the case of the second field in the quadratic estimator, normalized to sum to unity.
}
\label{fig:signalContribs}
\end{figure}

Optimal weight functions can easily be derived at lowest order by minimizing the Gaussian variance of the estimators~\cite{Hu:2001kj}.  In this paper we focus on the polarization quadratic estimators since polarization lensing reconstruction is a novel method which has not been investigated in detail for realistic applications. The full-sky
$EE$ and $EB$ quadratic estimators have response functions given by
\begin{equation}
\label{eq:fee}
f_{EE}(\vl_1,\vl_2)=\left[C_{l_1}^{EE}(\vL\cdot\vl_1)+C_{l_2}^{EE}(\vL\cdot\vl_2)\right]\cos(2\varphi_{\vl_1\vl_2})
\end{equation}
\begin{equation}
\label{eq:feb}
f_{EB}(\vl_1,\vl_2)=\left[C_{l_1}^{EE}(\vL\cdot\vl) - C_{l_2}^{BB}(\vL\cdot\vl_2)\right]\sin(2\varphi_{\vl_1\vl_2}),
\end{equation}
and the optimized weight functions are
\be
W_{EE}(\vl_1,\vl_2)=  \frac{f_{EE}(\vl_1,\vl_2)}{2\CEobs{l_1} \CEobs{l_2}},
\qquad
W_{EB}(\vl_1,\vl_2)=  \frac{f_{EB}(\vl_1,\vl_2)}{\CEobs{l_1} \CBobs{l_2}}.
\ee
 Here $\CEobs{l}$ and $\CBobs{l}$ power spectra are the observed $E$ and $B$ mode power spectra, the lensed power spectra plus instrument noise $\CEobs{l}=C_{l}^{EE}+\CEnoise_{l}$ and $\CBobs{l}=C_{l}^{BB}+\CBnoise_{l}$. The different trigonometric factors in the response function indicate that the contributions to the estimators come from rather different configurations: the $EE$ estimator has a lot of signal in squeezed shapes with $L\ll l_1,l_2$ and hence $l_1\sim l_2$, corresponding to reconstructing the large-scale lensing shear and convergence from the effect on the local small-scale power spectrum; however for the $EB$ estimator,  $\sin(2\varphi_{\vl_1\vl_2})\sim 0$ for $l_1\sim l_2$, and instead the dominant signal comes from correlating lensing-induced $B$ modes on a scale comparable to the lensing mode. This leads to a relatively modest reconstruction noise correlation between the estimators, especially on large scales, so the combination can significantly reduce the variance if the noise level is not low enough that the $EB$ mode estimator dominates (because there are no unlensed small-scale $B$ modes to contribute to the estimator variance). See Fig.~\ref{fig:signalContribs} for the contributions to the lensing signal at various different scales.

\subsection{Cut sky and $E$/$B$ leakage}
\label{sec:pureBintro}

The CMB $E$ and $B$ modes are defined as a harmonic transform of the $Q$ and $U$ Stokes parameters without a boundary. In the presence of a boundary (as on cut sky maps), the harmonics are no longer orthogonal, causing power to be leaked from the dominant $E$ mode into the subdominant $B$ mode if they are naively evaluated over only the observed patch of sky.  A number of methods have been developed to remove the spurious $B$ mode power originating from non-periodic boundary conditions on small patches of sky, e.g. \cite{Lewis:2001hp,Bunn:2002df,Smith:2005gi, Smith:2006vq}. A clean separation into pure-$B$ modes is effectively optimal for small noise levels where leakage from $E$ is dominating the variance of the contaminated observed $B$ modes. For intermediate noise levels inverse variance filtering would appropriately down weight the contaminated modes in an optimal way, and a full implementation of a nearly-optimal lensing reconstruction method~\cite{Hirata:2002jy,Hirata:2003ka,Hanson:2009gu} should therefore optimally handle the mixing effect at the expense of a very numerically costly inverse-variance filtering step.

In this paper we focus on suboptimal but simple methods for handling the cut sky as used by some recent ground-based observations, where a window function $W(\vx)$ is used to apodize the observed area smoothly to zero at the boundaries of the observed region. Pseudo harmonics are defined by directly transforming $W(\vx)(Q\pm iU)$:
\be
[\pseudoE \pm i\pseudoB](\vl) \equiv -\int d^{2}\vx W(\vx)[Q\pm iU](\vx) e^{\mp 2i\varphi_{l}}e^{-i\vl \cdot \vx},
\ee
however $\pseudoB$ will in general be a mixture of physical $E$ and $B$ modes since $e^{\mp 2i\varphi_{l}}e^{-i\vl \cdot \vx}$ are not orthogonal with respect to $W$.
Quantities that depend only on $E$ and $B$ modes can be obtained by choosing a general real window function $w$ that vanishes along with its derivative on the boundary of the observed region and outside. The quantities $E_w$ and $B_w$ defined by
\be
E_w \pm iB_w =   \int d^2\vx \,w (\partial_x \mp i \partial_y)^2 (Q \pm iU)   = \int d^2\vx (Q \pm iU) (\partial_x \mp i \partial_y)^2 w
\ee
are then guaranteed to be pure-$E$ and pure-$B$~\cite{Lewis:2001hp,Bunn:2002df}. An essentially optimal separation can be performed by using a complete basis of window functions, at the expense of considerable numerical cost. Alternatively, Ref.~\cite{Smith:2006vq} suggests a faster (but suboptimal) method using a set of pure modes obtained by taking $w = l^{-2}We^{-i\vl\cdot \vx}$, which reduces to the standard harmonics in the full sky case that $W=1$ everywhere. Since the small-scale $B$ mode signal is expected to be much smaller than the $E$ modes, the main concern is leakage of $E$ into $B$ rather than vice versa. We therefore only consider the pure-$B$ modes given for a particular choice of window $W(\vx)$ by
\be
B^{\rm pure}(\vl) \equiv \frac{1}{2l^2i}\int d^2\vx \left[(Q + iU) (\partial_x - i \partial_y)^2 -
(Q - iU) (\partial_x + i \partial_y)^2 \right](W e^{-i\vl\cdot \vx}) .
\ee
These modes are numerically simple to compute, and given explicitly by expanding the derivatives:
\begin{multline}
\label{Bpure_flat}
B^{\rm pure}(\vl)  = \int d^{2}\vx e^{-i\vl \cdot \vx } \Big(  \left[\sin(2\varphi_l)Q- \cos(2\varphi_l)U\right]W
  +\frac{2i}{l}\left[(Q\partial_y W - U \partial_x W)\cos\varphi_l + (U\partial_yW + Q\partial_x W)\sin\varphi_l\right]  \\
 +\frac{1}{l^{2}}\left[  U(\partial_x^2-\partial_y^2)W - 2Q\partial_x\partial_y W\right]\Big).
\end{multline}

The leakage of $E$ modes into $B$ modes is determined by the shape of the window function, with fractionally significant $B$ modes being generated from $E$ modes on a scale comparable to the variation of the window. To the extent that lensing reconstruction is using information only on scales much smaller than the variation of the window, one might expect the impact of the mixing to be modest. However as shown in Fig.~\ref{fig:signalContribs} the $EB$ polarization lensing reconstruction depends on $B$ modes from a very wide range of scales: even large-scale mixing effects are potentially important for $EB$-reconstruction, even though the $EE$ reconstruction information is mostly coming from very small scales.
In section~\ref{sec:cutsky} we compare the performance of the $EB\times EB$ and $EB\times EE$ lensing power spectrum estimators with and without the projection into pure-$B$ modes, to assess the importance of the mixing effect and the efficiency of using pure-$B$ modes to mitigate it in a straightforward way. Ref.~\cite{Smith:2006vq} also considers optimization of the window function, but here we just take the window to be a free function that we choose for convenience, so the results are expected to be slightly suboptimal.

\section{Polarization reconstruction without boundaries}
\label{sec:fullsky}

\begin{figure*}
\centerline{
\includegraphics[width=2.5in]{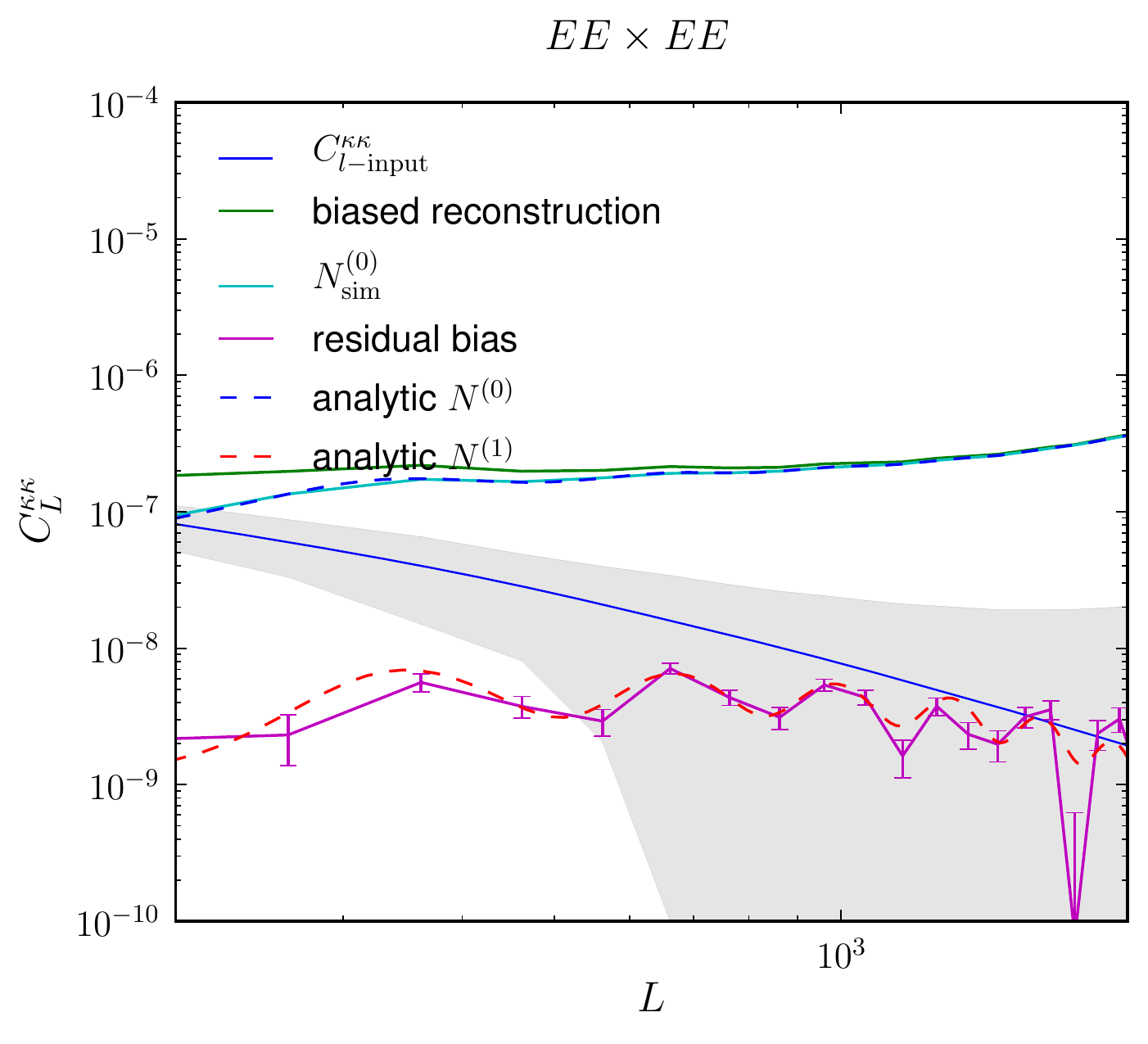}
\includegraphics[width=2.5in]{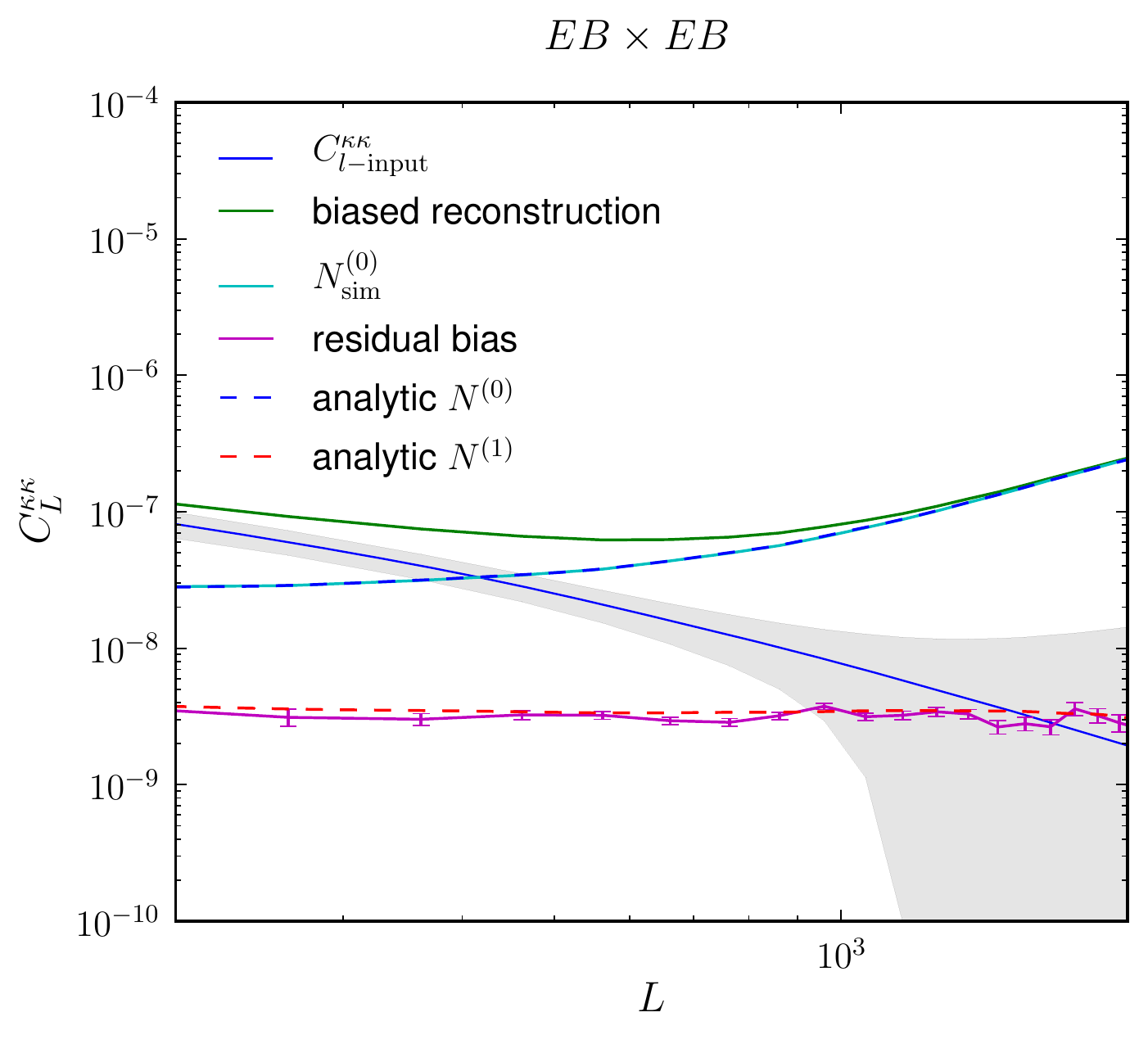}
\includegraphics[width=2.5in]{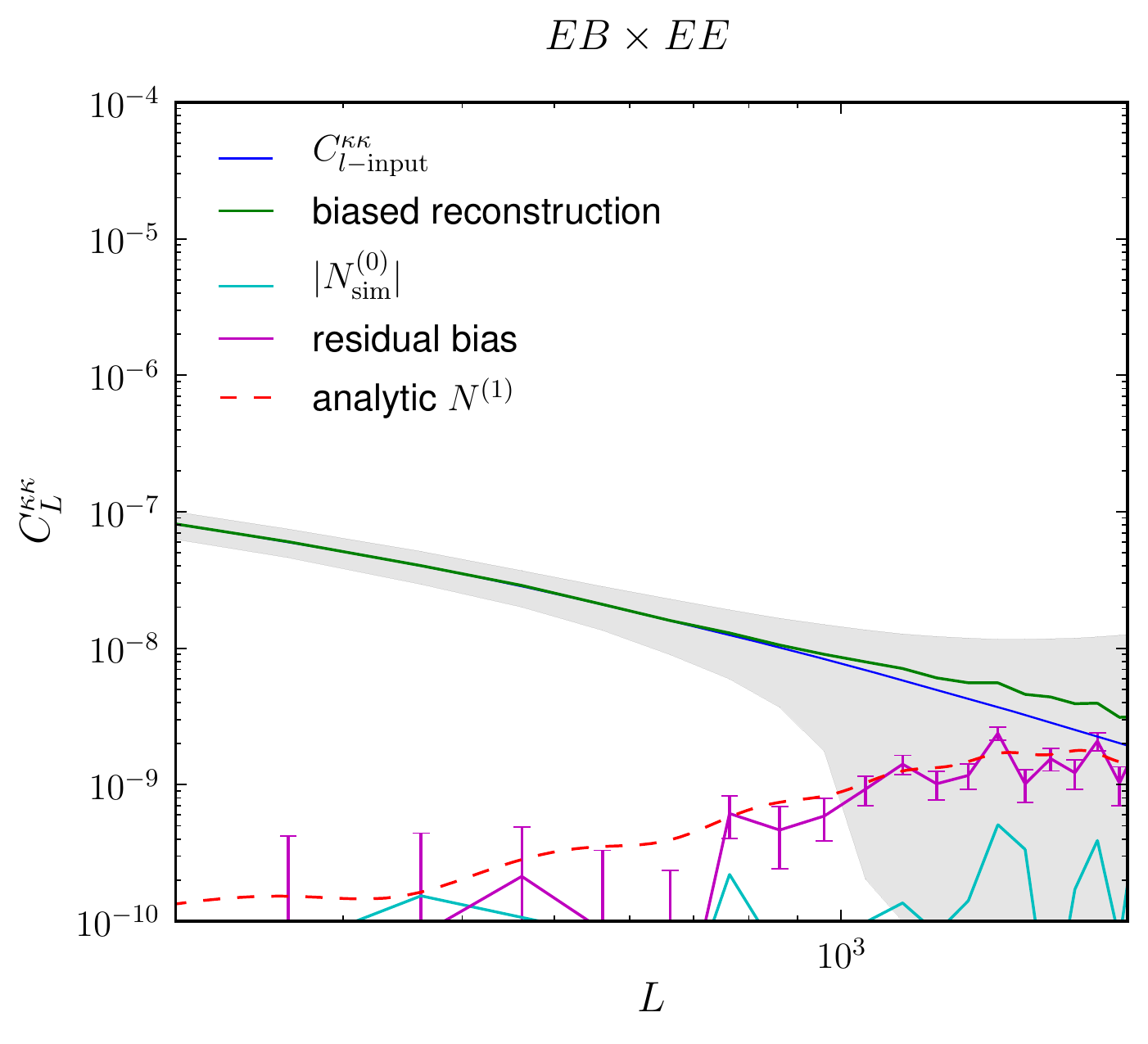}
}
\caption{
The lensing reconstruction power spectra from the $EE\times EE$, $EB\times EB$ and $EB\times EE$ estimators on a $9^{\circ}\times9^{\circ}$ patch of sky with periodic boundaries. Lines show the biased reconstruction, $\NZero_{\text{sim}}$, and residual bias from 1000 simulations, binned with $\Delta L=100$.  The binned one-sigma error on the reconstruction is shown by the grey band for any single realization, while the error bars on the residual bias are Monte Carlo errors from the simulations.  Analytic values of the $\NZero$ and $\NOne$ biases are also shown for comparison (dashed lines).
\label{fig:full}}
\end{figure*}

In this section we present results from simulations of the lensing reconstruction on a small patch of sky with periodic boundaries, so there are no issues of $E$/$B$ mixing.
Mock data CMB maps of the $Q$ and $U$ polarization are generated for a $9^{\circ}\times9^{\circ}$ patch of sky with a full width half maximum beam size of $\sigma=1.4 \text{arcmin}$.
Each unlensed simulation takes a random realization of theoretical unlensed input power spectra  $\tilde{C}_{l}^{EE}$, $\tilde{C}_{l}^{BB}$. These maps get lensed according to a random realization of a theoretical input power spectrum $\Cdinput_{L}$, where the unlensed power spectra and lensing power spectrum are calculated to linear order using \CAMB~\cite{Lewis:1999bs} for a $\Lambda$CDM cosmology.
To simulate the lensing, unlensed $Q$ and $U$ maps are first generated at two times the resolution of the final output lensed $Q$ and $U$ maps.  For each realization of the lensing potential, maps of the real-space $x$- and $y$-deflections are generated, and lensed maps are produced from the unlensed $Q$ and $U$ maps by bicubic interpolation of the values at the undeflected positions. This is sufficient for pixelization artefacts in polarization lensing reconstruction to be sub-percent. Random isotropic Gaussian beam-deconvolved detector noise is added with a power spectrum corresponding to $\bigtriangleup_{p}=4\muKarcmin$:
\begin{equation}
\label{}
\CEnoise_{l} = \CBnoise_{l}= \Cnoise_{l}=(\bigtriangleup_{p})^{2}e^{l(l+1)\sigma^{2}/8\ln 2}.
\end{equation}
 The beam-deconvolved lensed noisy maps are then used as the input for the quadratic estimator of the lensing potential, which initially returns a biased reconstruction with expectation $C_{L}^{dd}+\NZero(L)+\text{residual}$.  To calculate the $\NZero$ bias from the simulations $\NZero_{\text{sim}}$, we apply the quadratic estimator to unlensed maps generated with lensed power spectra (denoted $\bar{E}(\vl)$ and $\bar{B}(\vl)$):
\begin{equation}
\label{eq:N0EE}
\hd^{N0}_{EE}(\vL)=  \frac{A_{EE}(L)}{L} \int\frac{d^2\vl}{(2\pi)^{2}}\bar{E}(\vl)\bar{E}(\vl ')\frac{f_{EE}(\vl,\vl')} {2\CEobs{l} \CEobs{l'}}
\end{equation}

\begin{equation}
\label{eq:N0EB}
\hd^{N0}_{EB}(\vL)=\frac{A_{EB}(L)}{L} \int\frac{d^2\vl}{(2\pi)^{2}}\bar{E}(\vl)\bar{B}(\vl ')\frac{f_{EB}(\vl,\vl ')}{\CEobs{l}\CBobs{l'}}.
\end{equation}

To obtain the $\NZero_{\text{sim}}$ bias power spectrum, we take the power spectra of Eqs.~\eqref{eq:N0EE} and \eqref{eq:N0EB} averaged over 1000 simulations, which we can use as a check of the analytic result on the full sky. We do not use the realization-dependent $\NZero$ subtraction here, which may be significantly better for an actual data analysis where the theory and noise power spectra are uncertain, and reconstruction noise correlations would otherwise need to be modelled~\cite{Hanson:2010rp,Schmittfull:2013uea}.

From $\lensEstbiased_{ijpq,L}$, the raw power spectrum of the deflection angle quadratic estimators on the lensed maps, we define the residual bias $\hat{r}$(L) to be the difference from the input theoretical power spectrum after the Gaussian $\NZero_{\text{sim}}$ bias has been subtracted. This is expected to be approximately $\NOne(L)$, and is estimated from the simulations using

\be
\hat{r}_{ijpq} (L)= \frac{1}{\Nsim}\sum^{\Nsim}_{\text{k=1}} \left[\lensEstbiased_{ijpq,L}\right]_{k} - \NZero_{ijpq,\text{sim}}(L)-\Cdinput_L,
\ee

where $\Nsim=1000$ and $ij,pq \in EE,EB$.  Fig.~\ref{fig:full} shows the average lensing reconstructions for the $EE\times EE$, $EB\times EB$ and $EE \times EB$ power spectrum estimators, along with the $\NZero_{\text{sim}}$ and the residual bias as described above.  For comparison we show the expected analytic $\NZero$ and $\NOne$ biases as described in Sec.~\ref{sec:lensintro}.

The analytic $\NZero$ biases are consistent with the simulated $\NZero_{\rm sim}$ within binning for each estimator.   For the $EB\times EE$ reconstruction we show the absolute value of the $\NZero$ from simulation $|\NZero_{\rm sim}|$.  Although the theoretical $\NZero=0$ for the $EB\times EE$ power spectrum reconstruction, in practice we find $\NZero_{\text{sim}}\neq0$ at a level which is small and negligible for the total reconstruction, believed to be induced by pixelization.  The $\NZero_{\text{sim}}$ biases are shown and discussed in more detail in Fig.~\ref{fig:N0s} in a later section of the paper.

Furthermore, the analytic $\NOne$ bias is broadly consistent with the residual bias $\hat{r}(L)$ within the 1$\sigma$ error bars from 1000 simulations.  At the level of accuracy required the $\NOne$ bias therefore seems to be an adequate model for the residual bias for polarization reconstruction on small periodic patches of sky.  As a test that our pipeline is working correctly, we also calculated the cross-correlation power of each lensing realization map with the reconstructed lensing map, which agreed well with the input theoretical lensing power spectrum.

As an aside we note that the formulation of the $EB$ estimator given in Ref.~\cite{Kesden:2003cc} is slightly suboptimal, as it is derived with a constraint that the estimator is symmetric under interchange of $E$ and $B$.  A comparison of the  $\NZero$ bias for the $EB$ quadratic estimator given in Ref.~\cite{Kesden:2003cc} compared to the form given by Ref.~\cite{Hu:2001kj} shows that the estimator of Ref.~\cite{Kesden:2003cc} has $\sim25\%$ larger reconstruction noise than that of Ref.~\cite{Hu:2001kj} on scales $l\agt 2000$.   We use the estimators of Ref.~\cite{Hu:2001kj} (updated with lensed spectra in the weights as described in Sec.~\ref{sec:lensintro}), since they are perturbatively optimal on the full sky.

\section{Polarization reconstruction on the cut sky}
\label{sec:cutsky}

In this section we examine the more realistic case of lensing reconstruction when there is a boundary to the observed region.  We simulate $EE\times EE$, $EB\times EB$ and $EB\times EE$ lensing power spectrum reconstruction on a cut patch of sky, and then show the improvement in the reconstruction for $EB\times EB$ and $EB\times EE$ when the pure-$B$ mode estimator is used rather than windowing $Q$ and $U$ directly. The underlying quadratic estimators for the cut sky non-periodic boundary cases are:

\begin{equation}
\label{quadEEcut}
\hd^{\text{cut}}_{EE}(\vL)=  \frac{A_{EE}(L)}{L} \int\frac{d^2\vl_1}{(2\pi)^{2}}\pseudoE(\vl_1)\pseudoE(\vl_2)\frac{f_{EE}(\vl_1,\vl_2)} {2\CEobs{l_1} \CEobs{l_2}}
\end{equation}

\begin{equation}
\label{quadEBcut}
\hd^{\text{cut}}_{EB}(\vL)=\frac{A_{EB}(L)}{L} \int\frac{d^2\vl_1}{(2\pi)^{2}}\pseudoE(\vl_1)\pseudoB(\vl_2)\frac{f_{EB}(\vl_2,\vl_2)}{\CEobs{l_1}\CBobs{l_2}}
\end{equation}

\begin{equation}
\label{quadEBpure}
\hd^{\text{pure}}_{EB}(\vL)=\frac{A_{EB}(L)}{L} \int\frac{d^2\vl_1}{(2\pi)^{2}}\pseudoE(\vl_1)B^{\text{pure}}(\vl_2)\frac{f_{EB}(\vl_1,\vl_2)}{\CEobs{l_1}\CBobs{l_2}}.
\end{equation}

\begin{figure}[hb]
  \begin{minipage}[b]{0.5\linewidth}
    \centering
    \includegraphics[width=8.8cm]{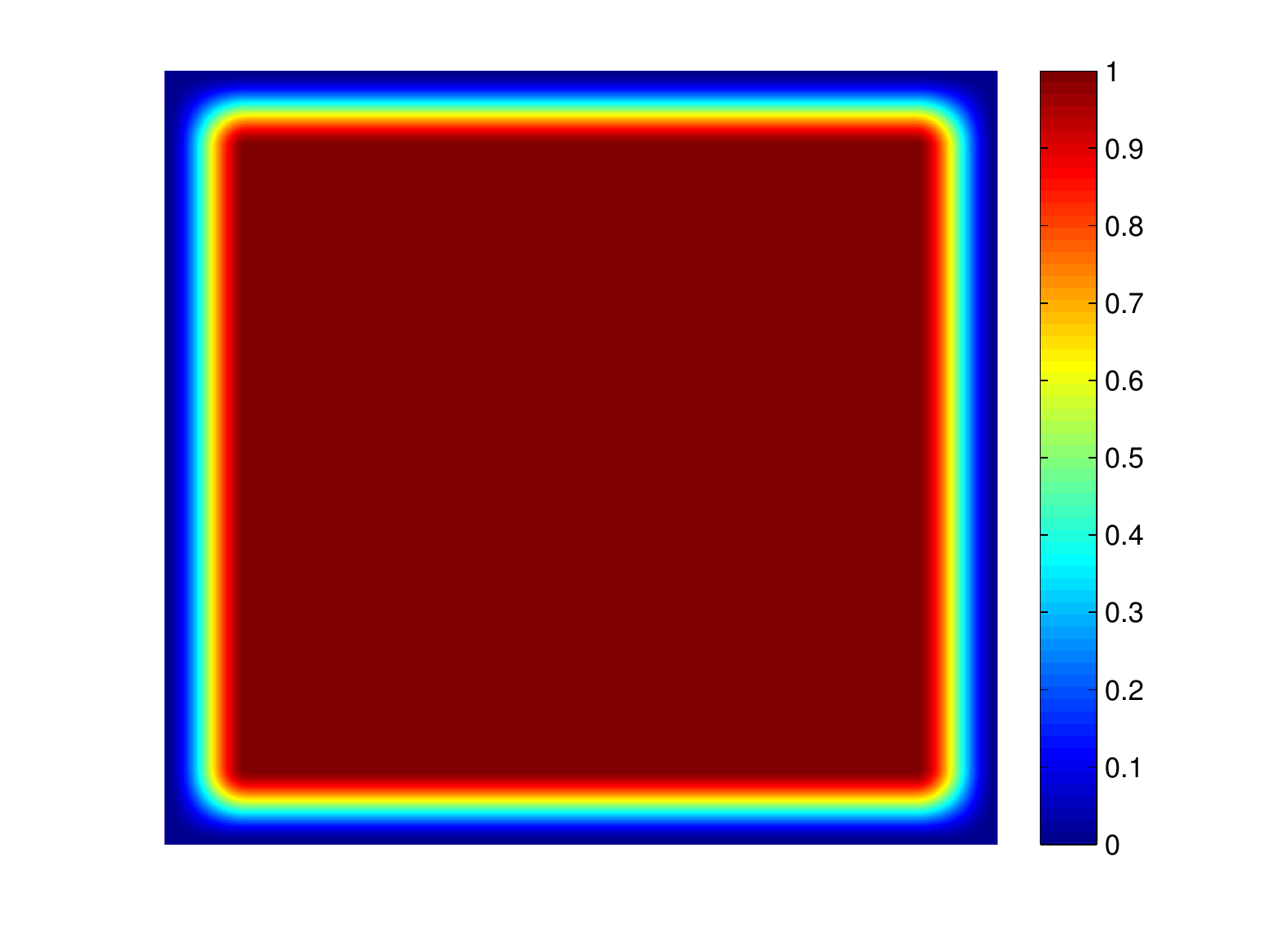}
     \caption{The apodization window}
           \label{fig:win}
  \end{minipage}
  \begin{minipage}[b]{0.5\linewidth}
    \centering
    \includegraphics[width=8.8cm]{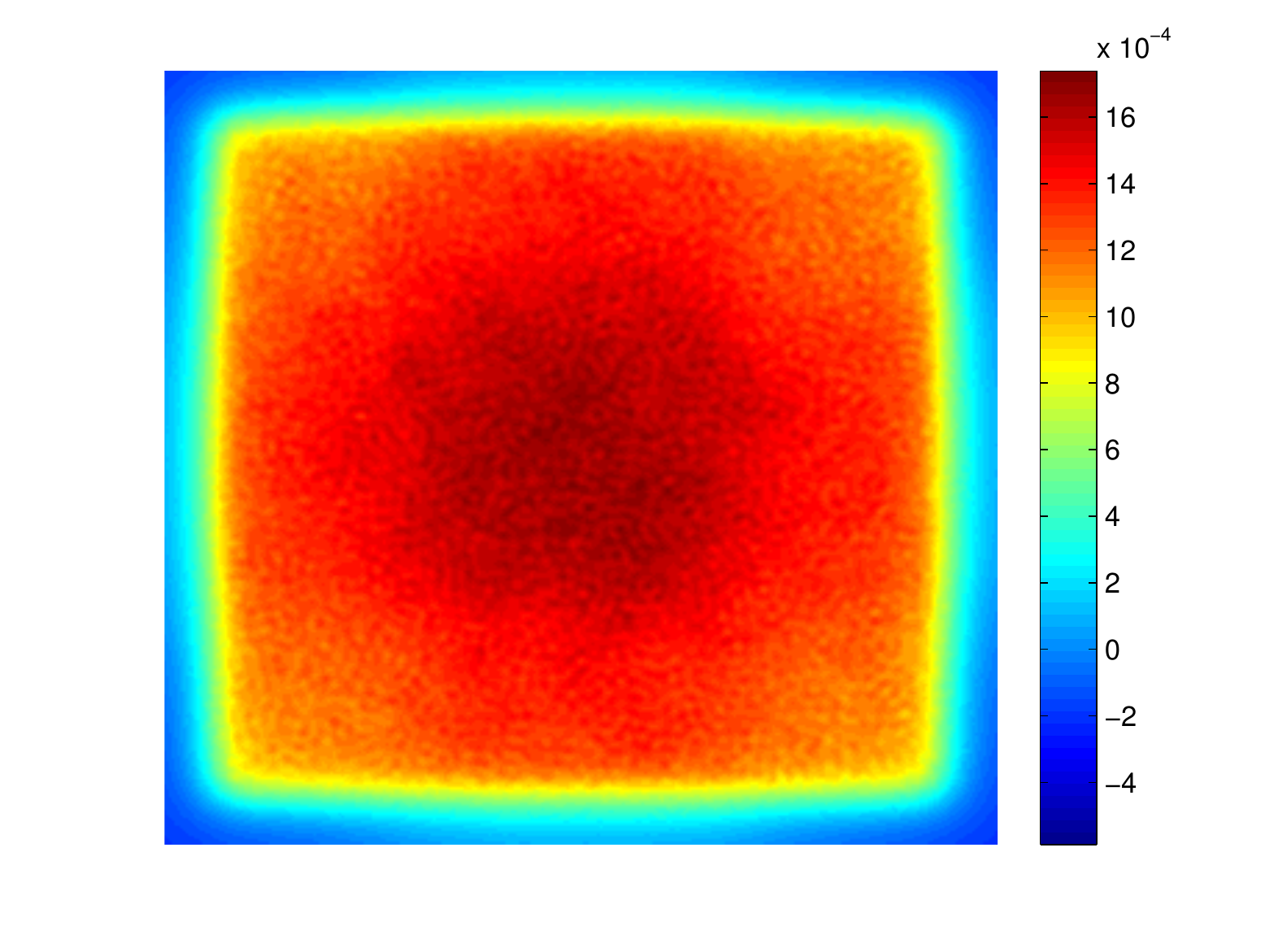}
    \caption{The mean field map for $\hd^{\rm cut}_{EE}$}
     \label{fig:MF_EE_cut}
  \end{minipage}
  \begin{minipage}[b]{0.5\linewidth}
    \centering
    \includegraphics[width=8.8cm]{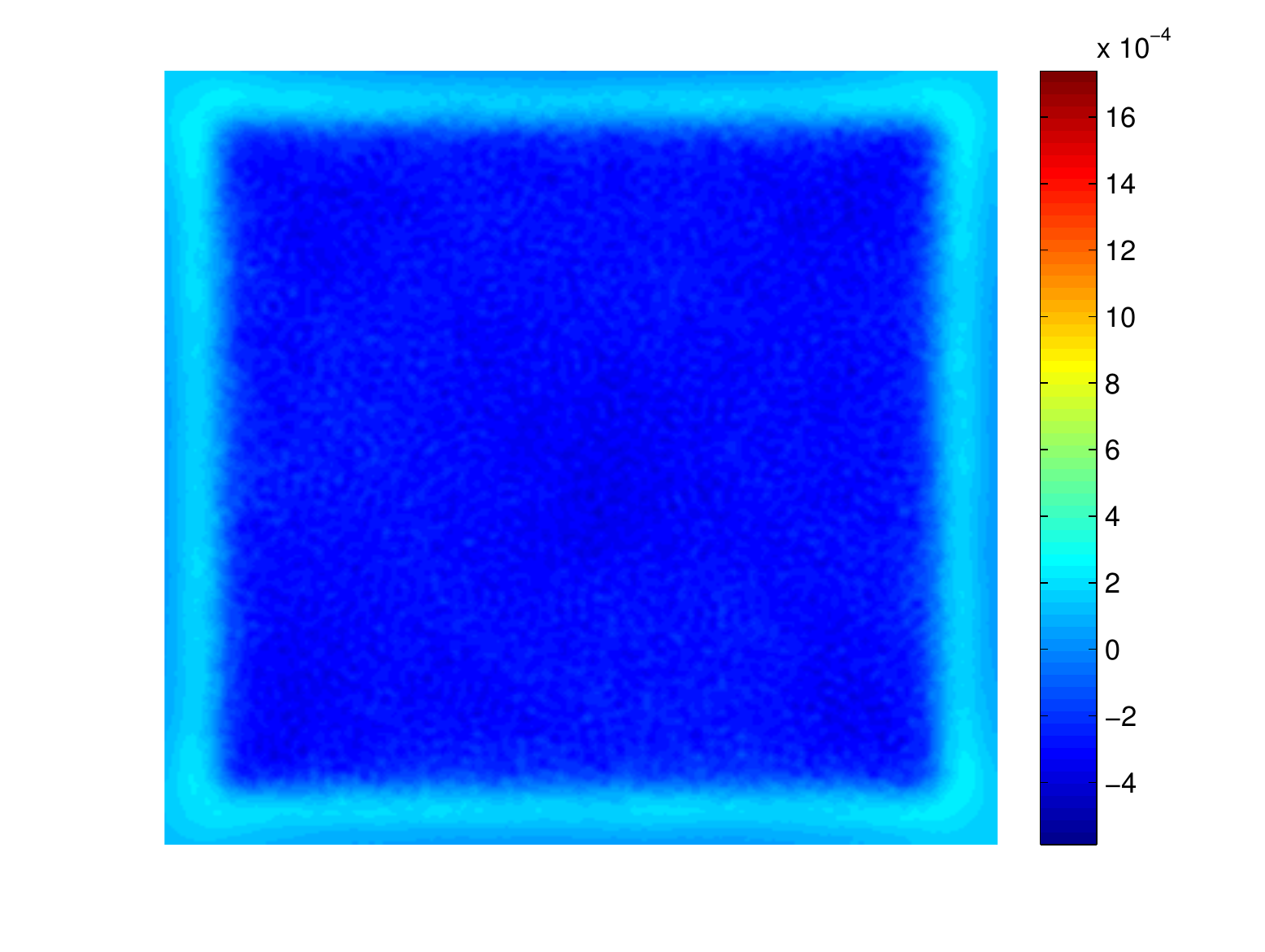}
    \caption{The mean field map for $\hd^{\rm cut}_{EB}$}
     \label{fig:MF_EB_cut}
  \end{minipage}
  \begin{minipage}[b]{0.5\linewidth}
    \centering
    \includegraphics[width=8.8cm]{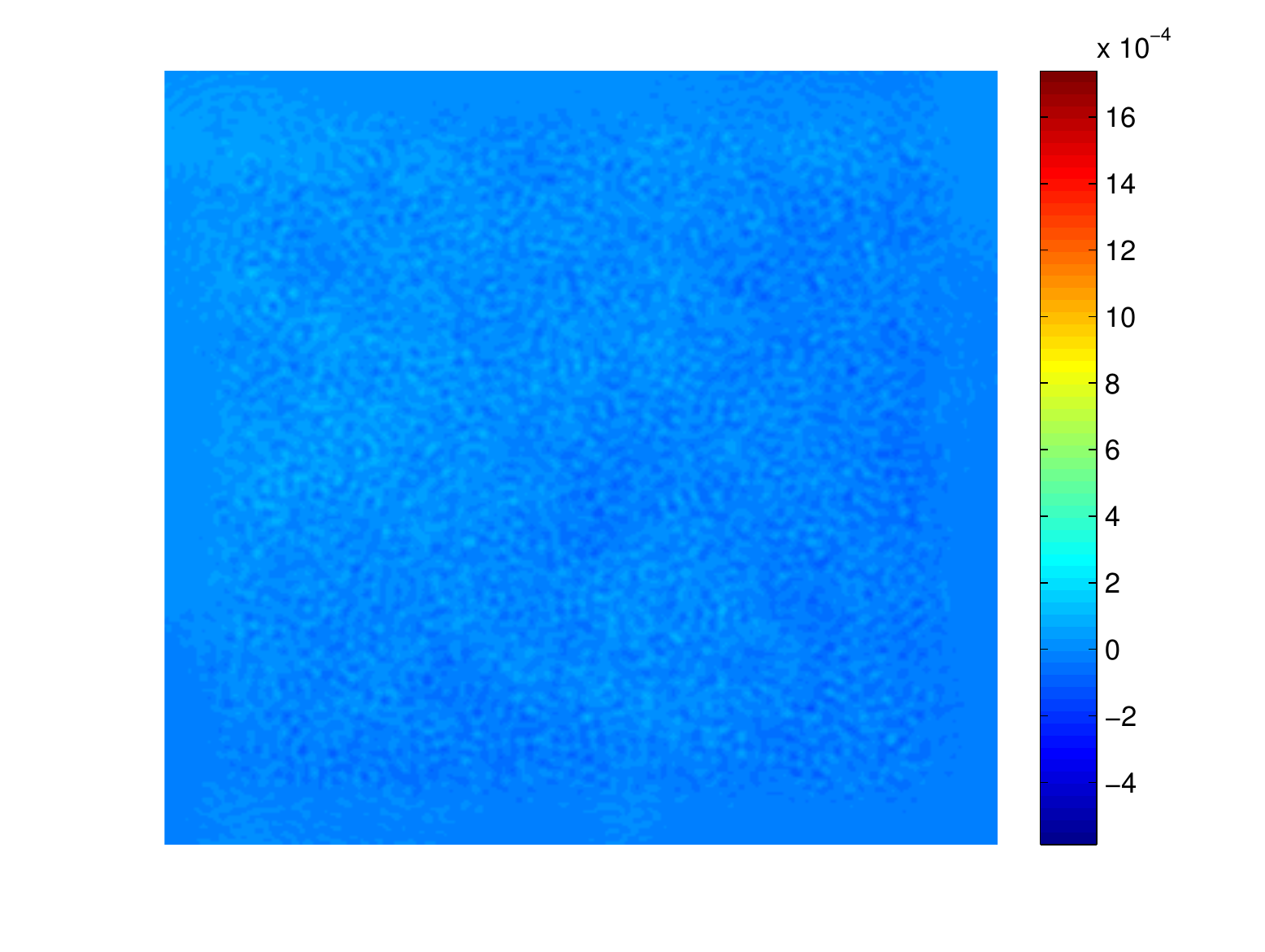}
    \caption{The mean field map for $\hd^{\rm pure}_{EB}$}
     \label{fig:MF_EB_pure}
  \end{minipage}
\end{figure}

 To make our simulated maps non-periodic, we cut out and use one quarter of the $9^{\circ}\times9^{\circ}$ periodic map, which is a $4.5^{\circ}\times4.5^{\circ}$ patch (now with non-periodic boundaries).  All other properties of the map simulation are unchanged from those described in section~\ref{sec:fullsky}.

To mitigate the effect of harmonic ringing, we use an apodization window which goes smoothly to zero at the edges, as required to construct the pure-$B$ mode estimator of Eq.~\eqref{Bpure_flat}.
 We use a window which contains mostly unit values except for a simple cosine tapering on the edge which is one tenth the size of the cut patch.  The window is shown in Fig.~\ref{fig:win}.  The tapering around the edge is a quarter-period cosine which is normalized such that the tapering falls smoothly from unity in the central area to zero at the map boundary over the one tenth edge. The cut sky patch consists of $600 \times 600$ pixels. As the deflection field is generated on the larger $9^{\circ}\times9^{\circ}$ patch before being cut, the cut patch contains modes down to $l_{\rm min}=20$ (although after cutting the angular scale of the patch is $l=40$).

 The cut sky and window introduce statistical anisotropy in the map, which gives rise to a spurious signal in the lensing reconstruction from $W(\vx)(Q+iU)(\vx)$. The average map-level bias is called the mean field~\cite{Hanson:2009gu}, and typically closely follows the shape of the window that is causing it. There can also be other sources of mean field, like beam asymmetries and anisotropic noise, but for simplicity we restrict our analysis to isotropic noise.
  For an ideal full-sky measurement the lensing estimators should average to zero, i.e. $\la \hd \ra =\la d \ra =0$, but this is no longer the case in the presence of a window. However, simulations can be used to estimate the mean field $\la \hd \ra$, which can then be subtracted from the lensing estimator to form an unbiased reconstruction $\hd - \la \hd \ra$.
  To obtain the mean field 1000 reconstructed lensing potential maps were averaged in the map space.  This mean field map was then removed from the reconstructed lensing potential maps prior to taking their power spectra.

\begin{figure*}
\centerline{
\includegraphics[width=4in]{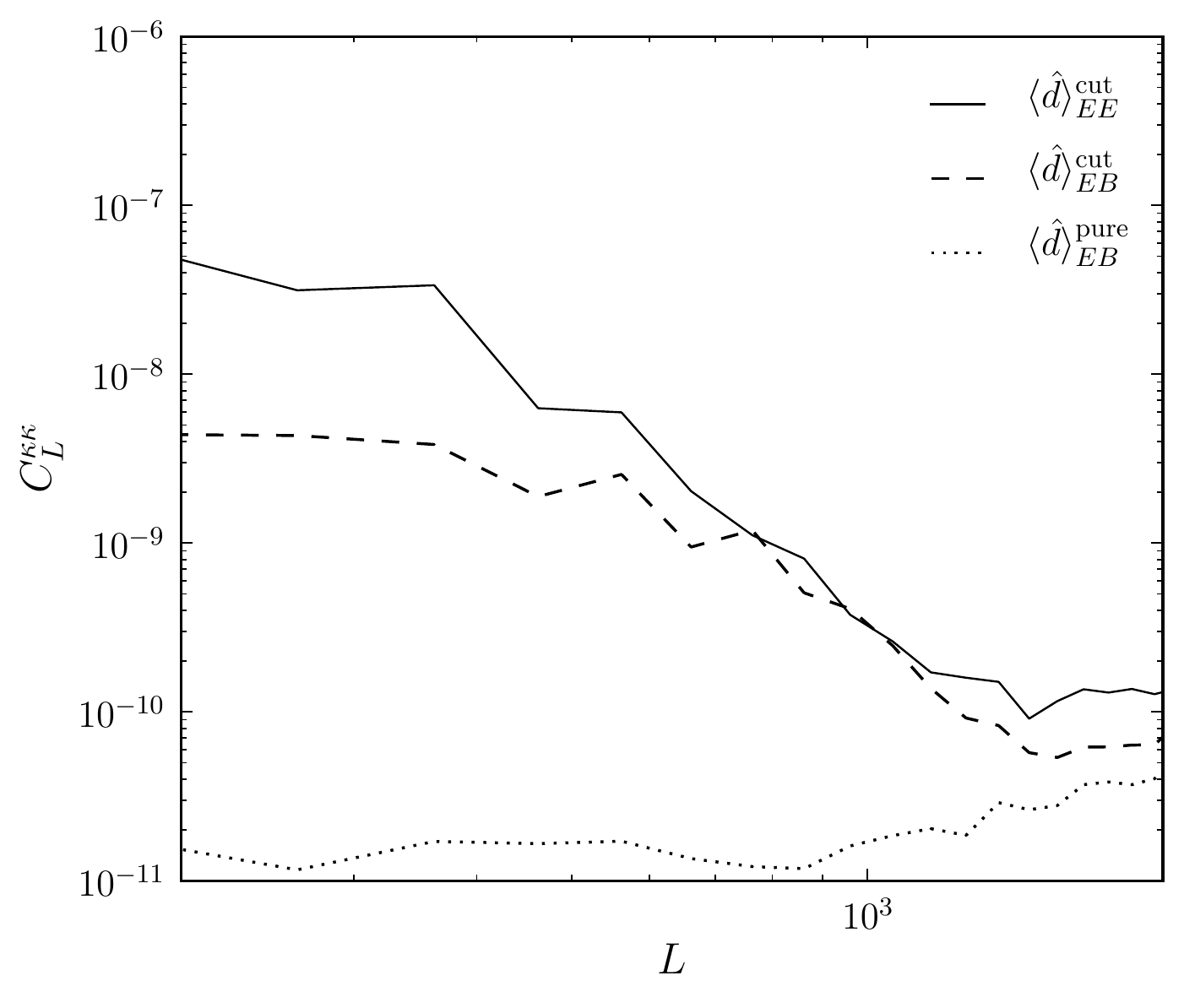}
}
\caption{
Power spectra of the mean field maps shown in Figs.~\ref{fig:MF_EE_cut}, \ref{fig:MF_EB_cut} and \ref{fig:MF_EB_pure}. The $EB$ mean field is substantially reduced by using pure-$B$ modes in the estimator, in good agreement with the low level of $EB$ mean field found by Ref.~\cite{Namikawa:2013xka} using inverse variance weighting.
}
\label{fig:residual_MFpow}
\end{figure*}

 Figs.~\ref{fig:MF_EE_cut}, \ref{fig:MF_EB_cut} and \ref{fig:MF_EB_pure} show the mean field maps from the three reconstruction estimators that we simulate on the cut sky: an $EE$ reconstruction, an $EB$ reconstruction using the cut sky $B$ modes, and an $EB$ reconstruction using pure-$B$ modes.  The corresponding mean-field power spectra are shown in Fig.~\ref{fig:residual_MFpow}.  Unsurprisingly, the mean field follows the shape of the window and is largest in the centre. The $EB$ mean field calculated using the pure-$B$ modes is much less than the $EE$ mean field, in agreement with Ref.~\cite{Namikawa:2013xka}. Without the projection into pure-$B$ modes this is no longer true, and the mean field is substantially larger.

The cut-sky estimators have a reduced amplitude due to the window having $W(\vx)<1$, and this must be accounted for when estimating power spectra from the windowed sky.

In the limit that all the modes of interest ($Q$, $U$ and lensing potential) are much smaller than the scale of variation of the window, the window can locally be regarded as a constant, and the lensing reconstruction estimator (which depends quadratically on $Q$ and $U$) is therefore simply the full-sky value multiplied by $[W(x)]^2$. The local power spectrum then scales like $[W(\vx)]^4$, and the value estimated over the full patch is therefore expected to be scaled by the average value of $W^4$. Likewise the variance of the power spectrum locally scales like $[W(\vx)]^8$.
Following Ref.~\cite{BenoitLevy:2013bc} we therefore use averaged $f_{W4}$ and $f_{W8}$ factors to account for the window in the power spectra and variance respectively, where on the pixelized map
\be
f_{Wn} \equiv \frac{1}{N_{\rm pix}} \sum_i \left[W(\vx_i)\right]^n.
\ee
 This is expected to be accurate for the intermediate-scale reconstruction from $EE$ (where all the information is in small-scale $E$ modes), but may be less accurate for the $EB$ reconstruction where the $B$-mode contribution is much less local. It is also likely to be inaccurate on large scales
 (comparable to the scale of variation of the window).

  All our simulated cut-sky power spectra ($\NZero_{\text{sim}}$, residual bias, biased reconstruction and unbiased reconstruction) have been scaled to account for the smaller sky fraction and the effect of the window via a scaling factor $f_{W4}$.

For example our power spectrum estimators for the cut sky are

\be
\lensEstbiased_{ijpq,L} \equiv \frac{1}{f_{W4}}\left[\lensEstcutbiased_{ijpq,L} \right].
\ee
For a periodic sky patch, neglecting first order ($\NOne$) biases, the approximate error in the lensing potential for the $EE\times EE$ and $EB\times EB$ power spectrum estimators is~\cite{Hu:2001kj}:
\begin{equation}
\label{ }
\triangle C_{L}^{dd}\approx \frac{1}{\sqrt{L\triangle L \fsky}}\left[C_{L}^{dd}+\NZero(L)\right],
\end{equation}
and for the $EB\times EE$ power spectrum estimator:
\begin{equation}
\label{ }
\triangle C_{L}^{dd}\approx \frac{1}{\sqrt{2L\triangle L \fsky}}\sqrt{ \left[C_{L}^{dd}+\NZero(L)\right]_{EE} \left[C_{L}^{dd}+\NZero(L)\right]_{EB} + \left(C_{L}^{dd}\right)^{2} },
\end{equation}
where $\Delta L$ is the bin size.  For a windowed sky patch, the error in a measurement of the lensing potential is modified to approximately~\cite{BenoitLevy:2013bc}:
\begin{equation}
\label{eq:varcut}
\triangle C_{L}^{dd-\text{cut}}\approx \sqrt{\frac{f_{W8}}{f^{2}_{W4}}} \triangle C_{L}^{dd},
\end{equation}
if there are no issues of $E$/$B$ mixing. On small scales with lower noise, the error bars would be significantly increased due to $\NOne$ biases which couple in cosmic variance from larger-scale modes.

\begin{figure*}
\centerline{
\includegraphics[width=2.5in]{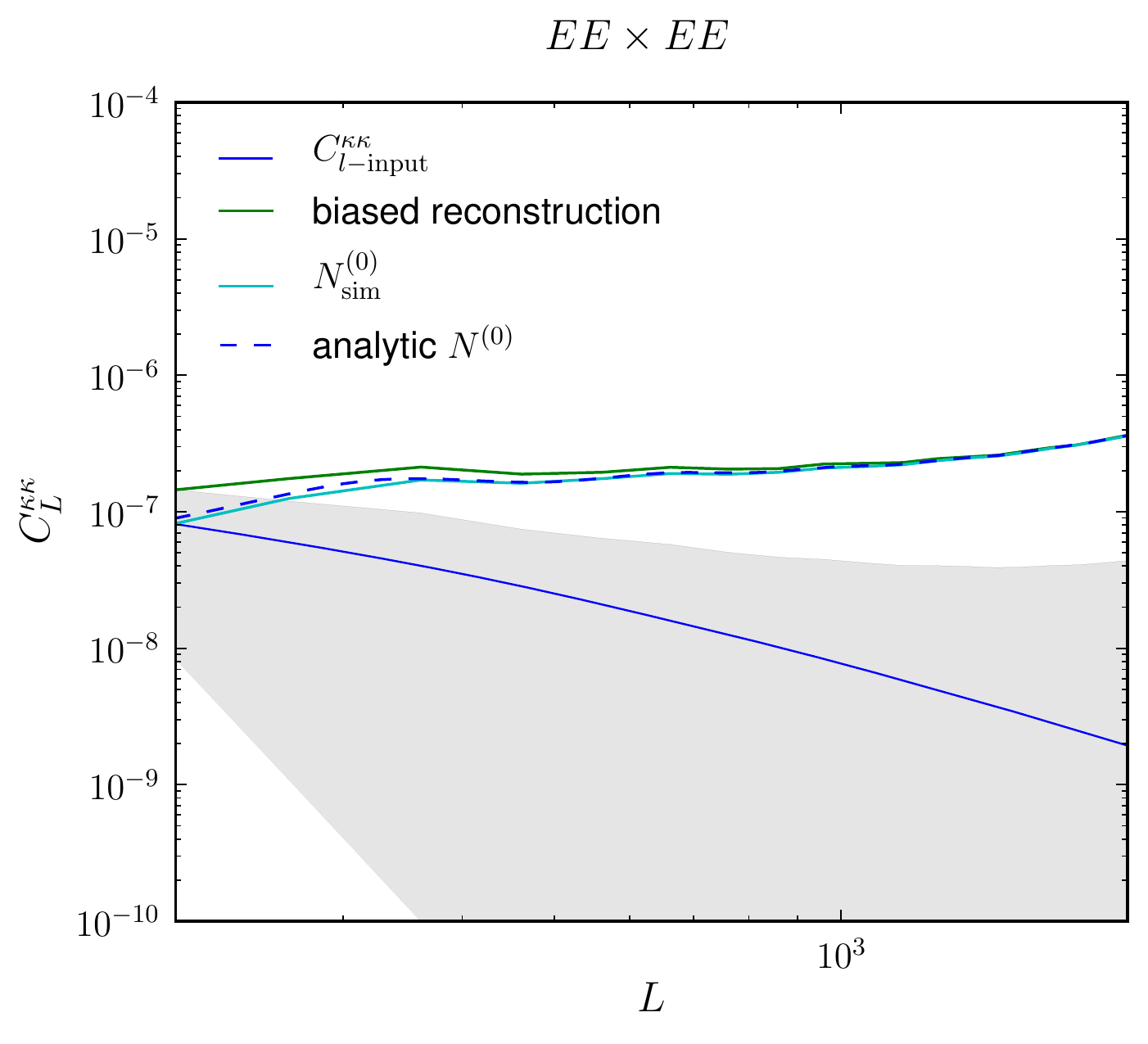}
\includegraphics[width=2.5in]{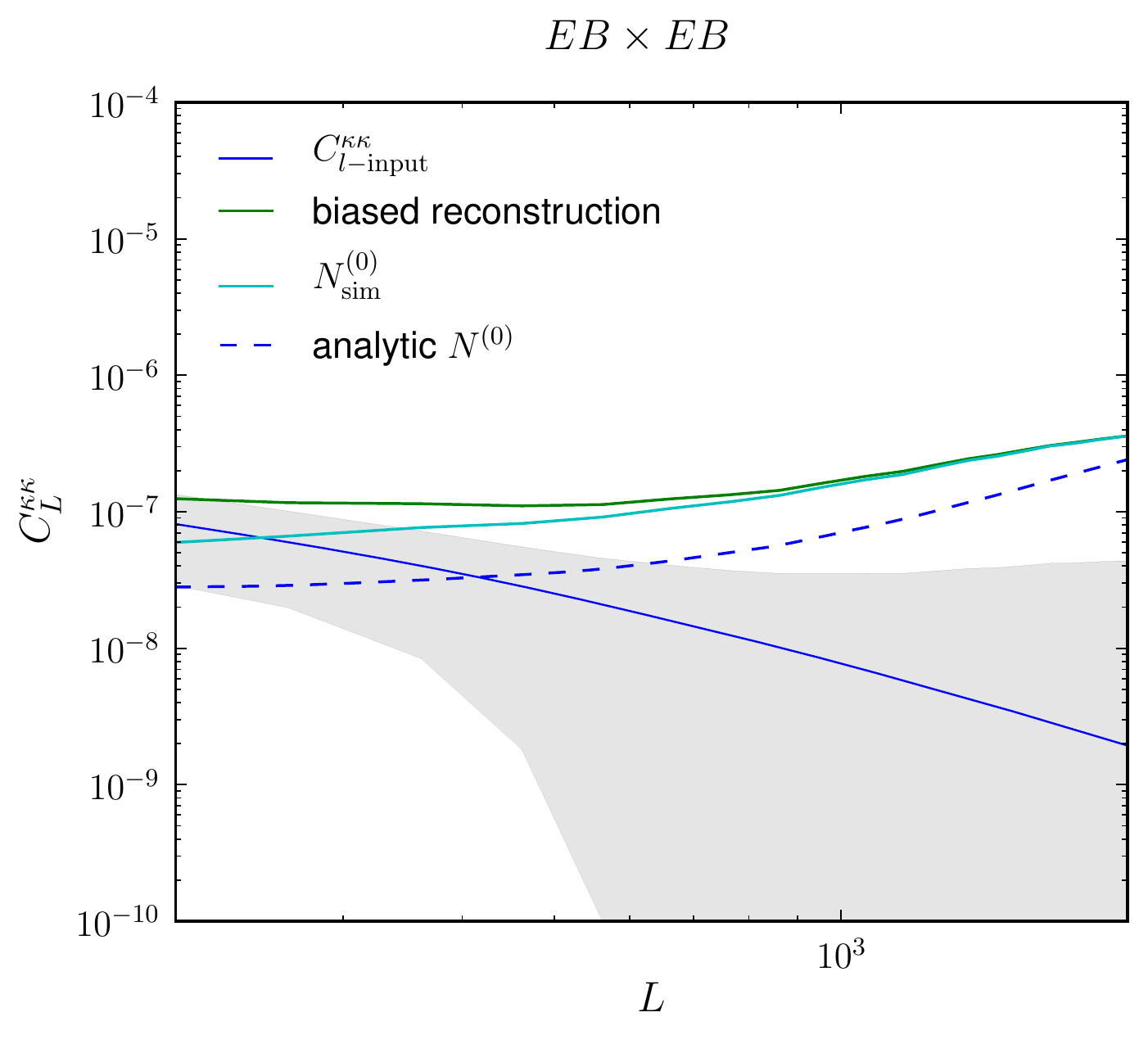}
\includegraphics[width=2.5in]{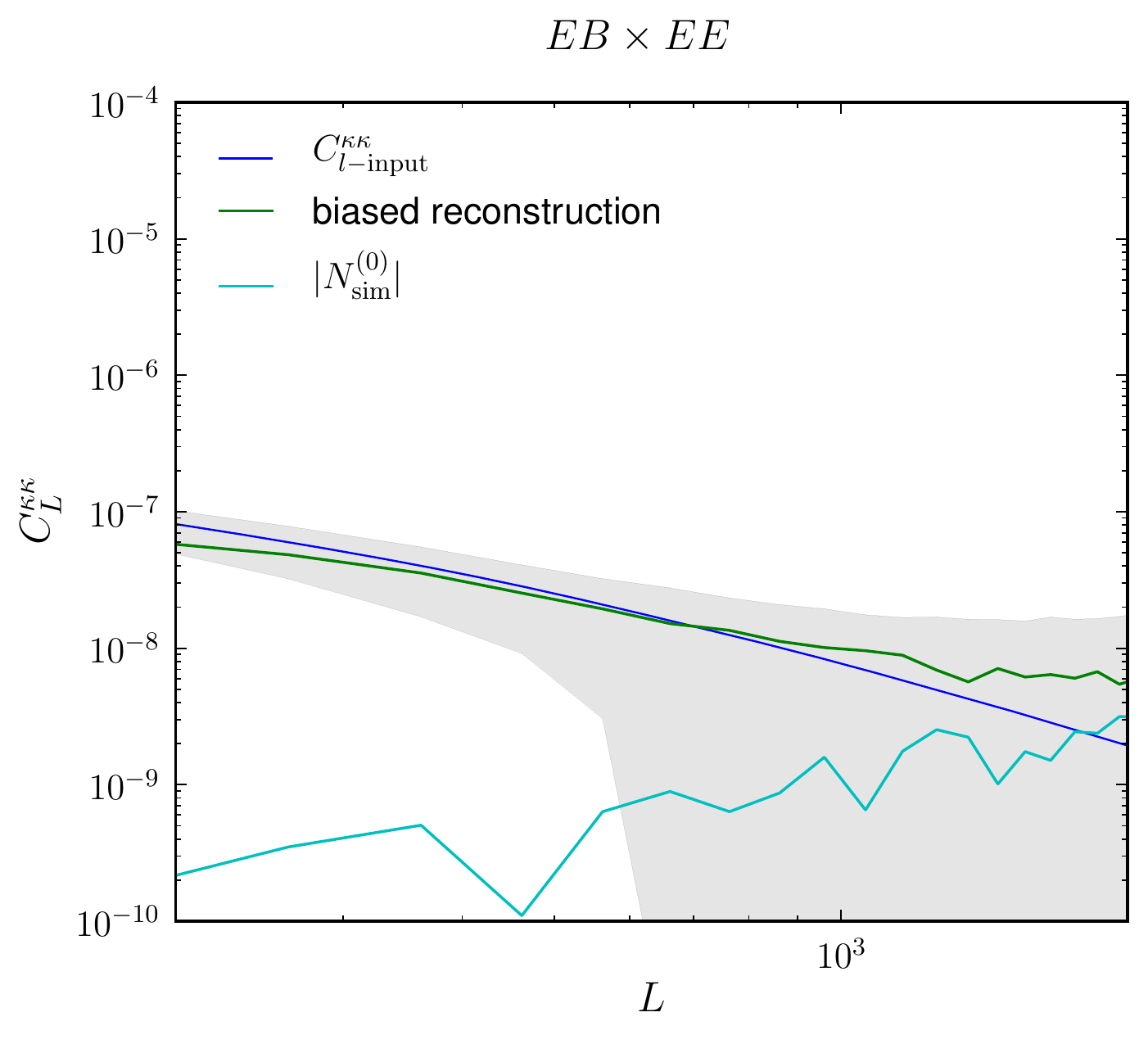}
}
\caption{
The lensing reconstruction using the $EE\times EE$, $EB\times EB$ and $EB\times EE$ power spectrum estimators on a $4.5^{\circ}\times4.5^{\circ}$ apodized cut patch of sky (without $E$/$B$ separation).  The binned one sigma error on the reconstruction is shown by the grey band for any single realization.  The full-sky analytic $\NZero$ bias is also shown for comparison.  Results shown are from 1000 simulations.
}
\label{fig:cut}
\end{figure*}

\begin{figure*}
\centerline{
\includegraphics[width=3.5in]{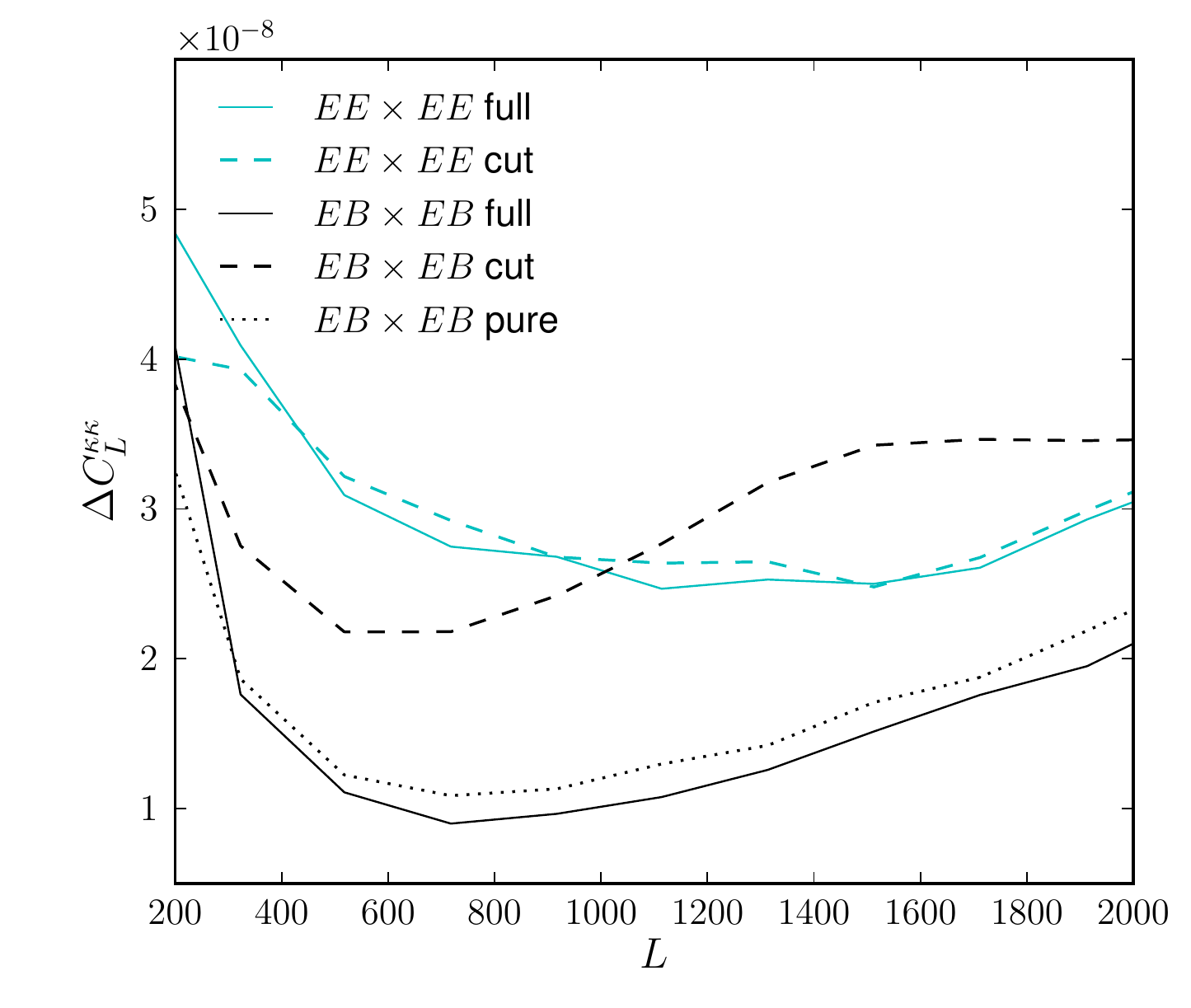}
\includegraphics[width=3.5in]{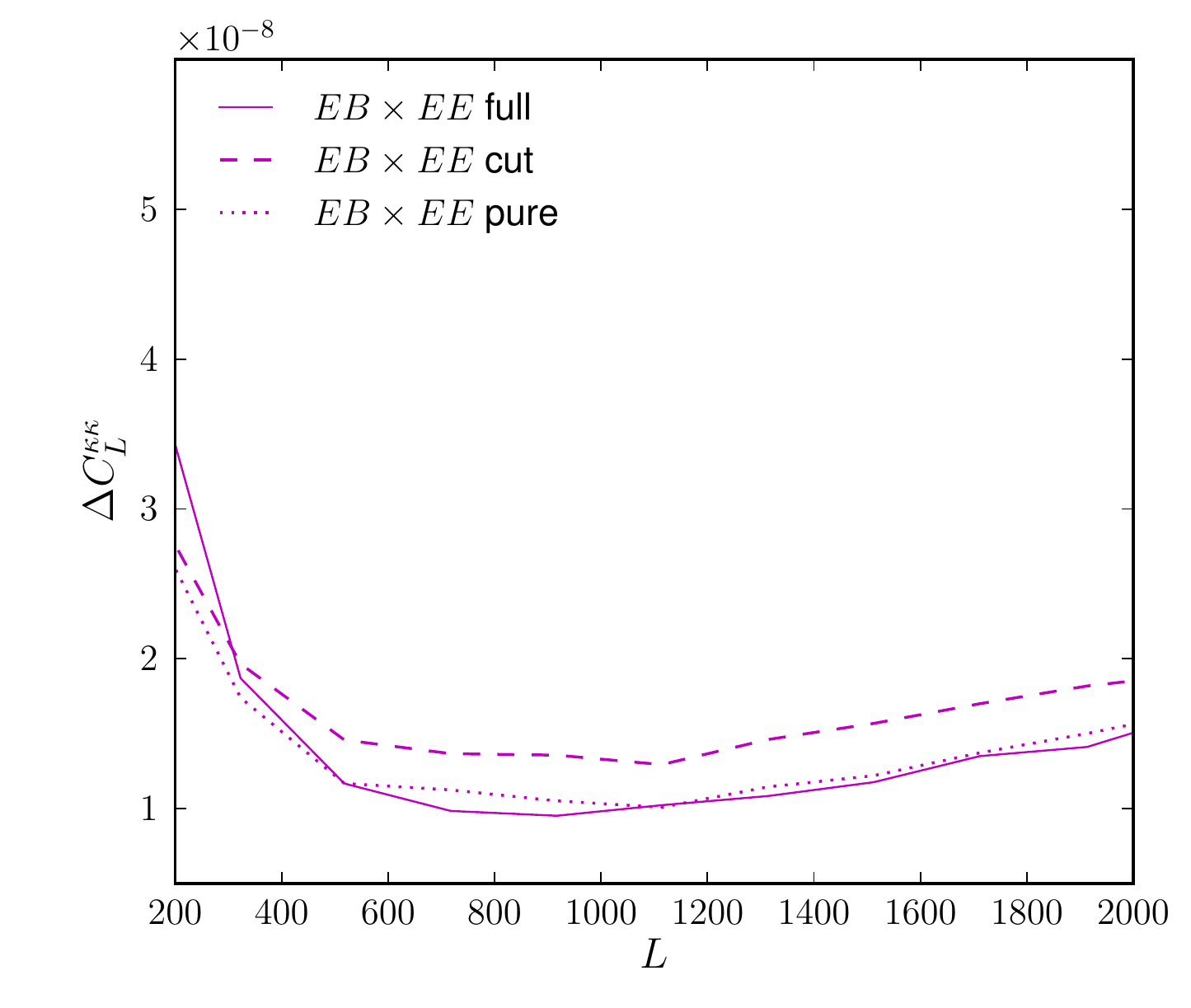}
}
\caption{
A comparison of the lensing power spectrum reconstruction errors, which are significantly reduced by using pure-$B$ rather than cut-sky $B$ modes in the estimators involving $B$ modes.  Note that error bars on the periodic sky have been scaled to have the same sky fraction as a periodic $4.5^{\circ}\times4.5^{\circ}$ patch, and the binning here is $\Delta L=200$ to reduce correlation between bins.  Left:  The $EE \times EE$ and $EB\times EB$ reconstruction error bars.  Right:  The $EB \times EE$ reconstruction error bars.
}
\label{fig:errorbar}
\end{figure*}

For our choice of window function the scaling factors are $f_{W4}=0.1826, f_{W8}=0.1761$, which are less than the $0.25$ value one would get from a quarter patch without apodization. We also use a top-hat binning with size $\Delta L=100$.  Using a bin size of e.g. $\Delta L=50$, close to the window scale, led to correlations between the bins, causing the variance to be underestimated when not accounting for covariances. Tests showed that $\Delta L=100$ is large enough to prevent large correlations between bins, although for the $\fsky$-scaled comparison of the error bars shown in Fig.~\ref{fig:errorbar} we use $\Delta L=200$ to reduce correlations to a lower level.  The cut-sky reconstructed power spectra for $EE\times EE$, $EB\times EB$ and $EB\times EE$ are shown in Fig.~\ref{fig:cut}.

\begin{figure*}
\centerline{
\includegraphics[width=3.5in]{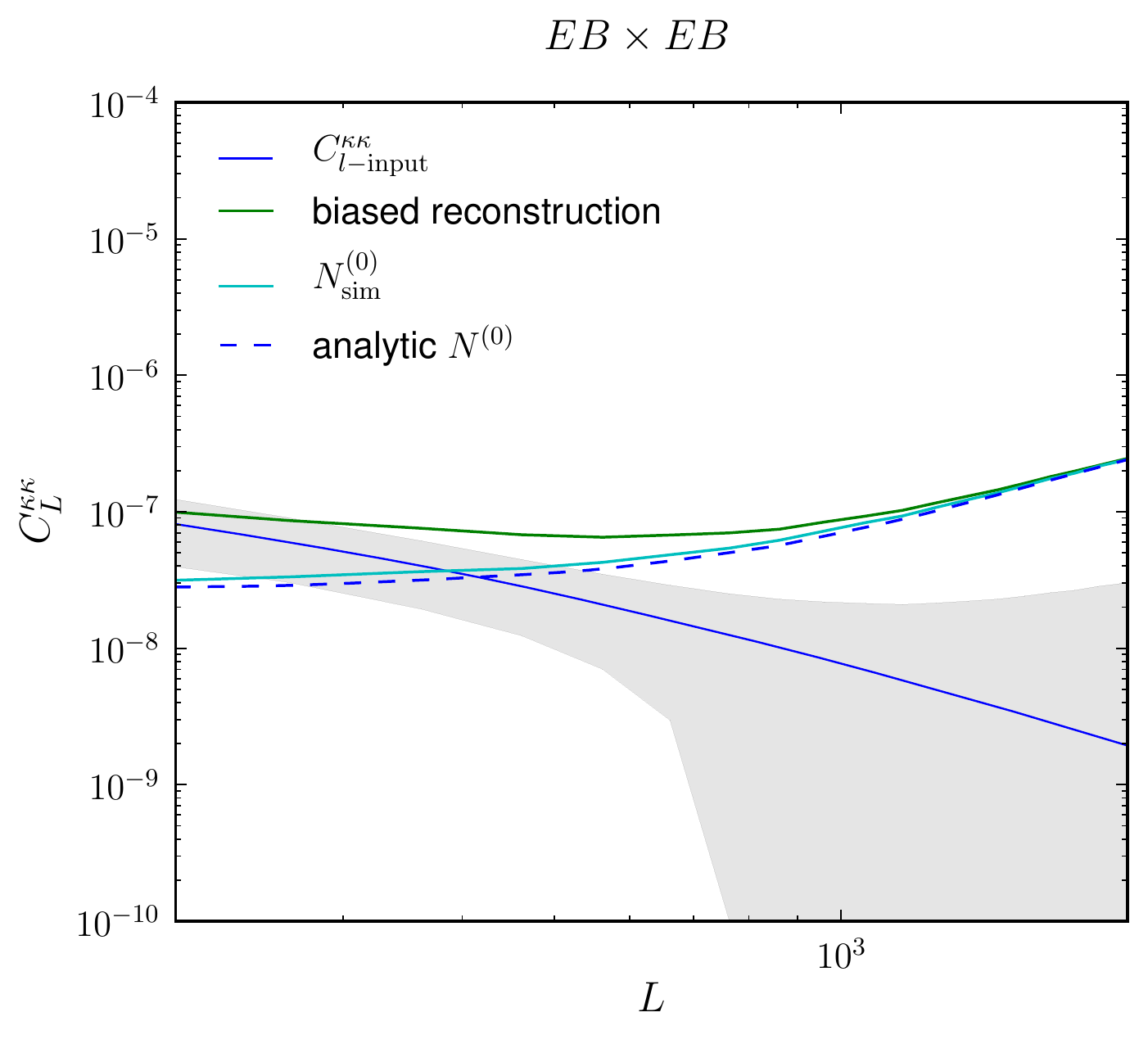}
\includegraphics[width=3.5in]{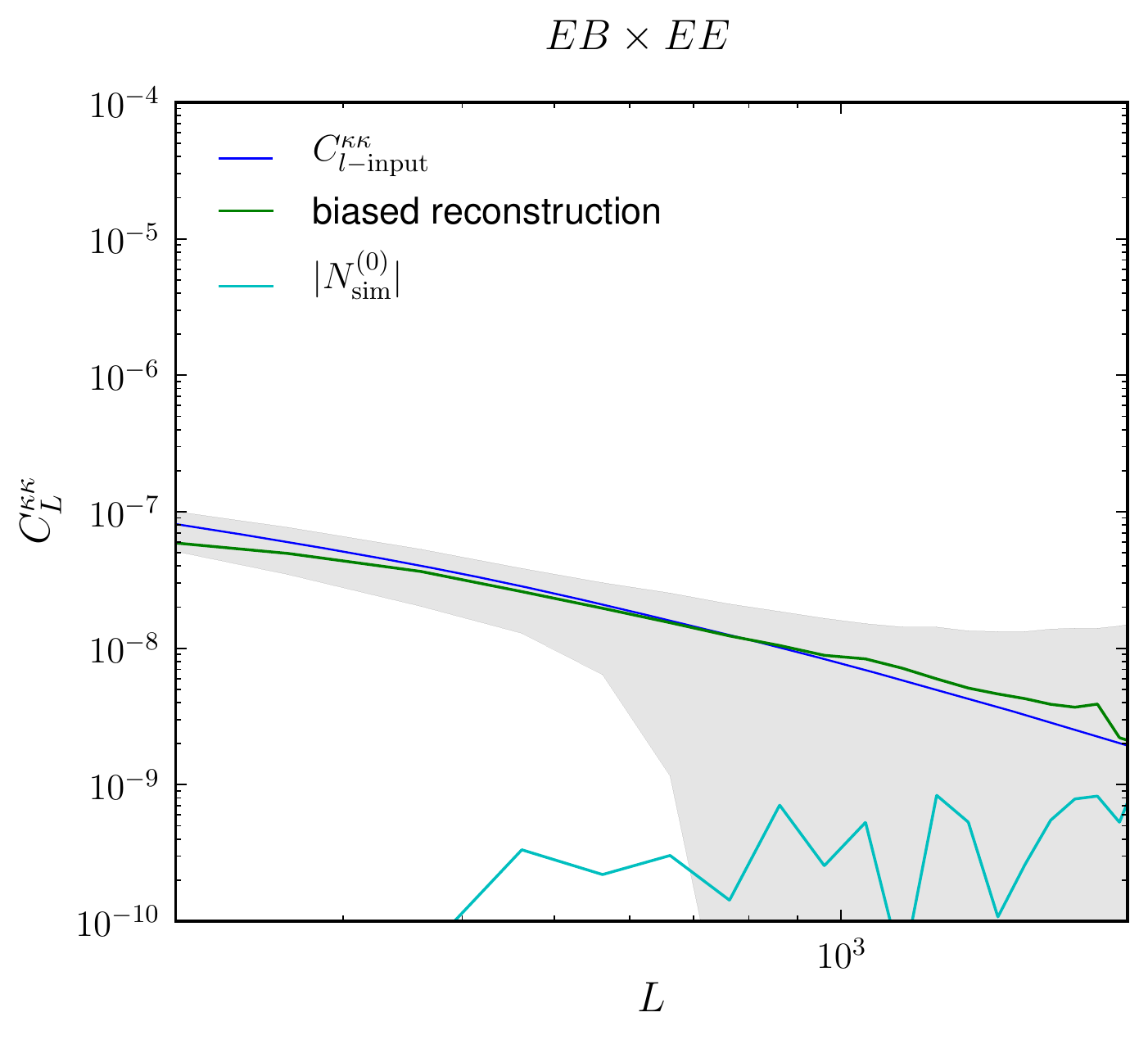}
}
\caption{
The lensing reconstruction using the $EB\times EB$ and $EB\times EE$ quadratic estimators as in Fig.~\ref{fig:cut}, but now using pure-$B$ modes in the estimators.
}
\label{fig:pure}
\end{figure*}

In all the cut-sky reconstructions, the error bars are larger than the reconstructions without boundaries, as expected due to the significantly reduced effective area.  The leakage of $E$ into $B$ modes is also expected to increase the non-lensing $B$ mode power, and hence increase error in the reconstructions involving cut-sky $B$ modes. Removing this leakage should reduce the error bars. In Fig.\ref{fig:pure} we show that the errors are indeed significantly reduced by using pure-$B$ modes in the $EB\times EB$ and $EB\times EE$ estimators.

\begin{figure*}
\centerline{
\includegraphics[width=3.5in]{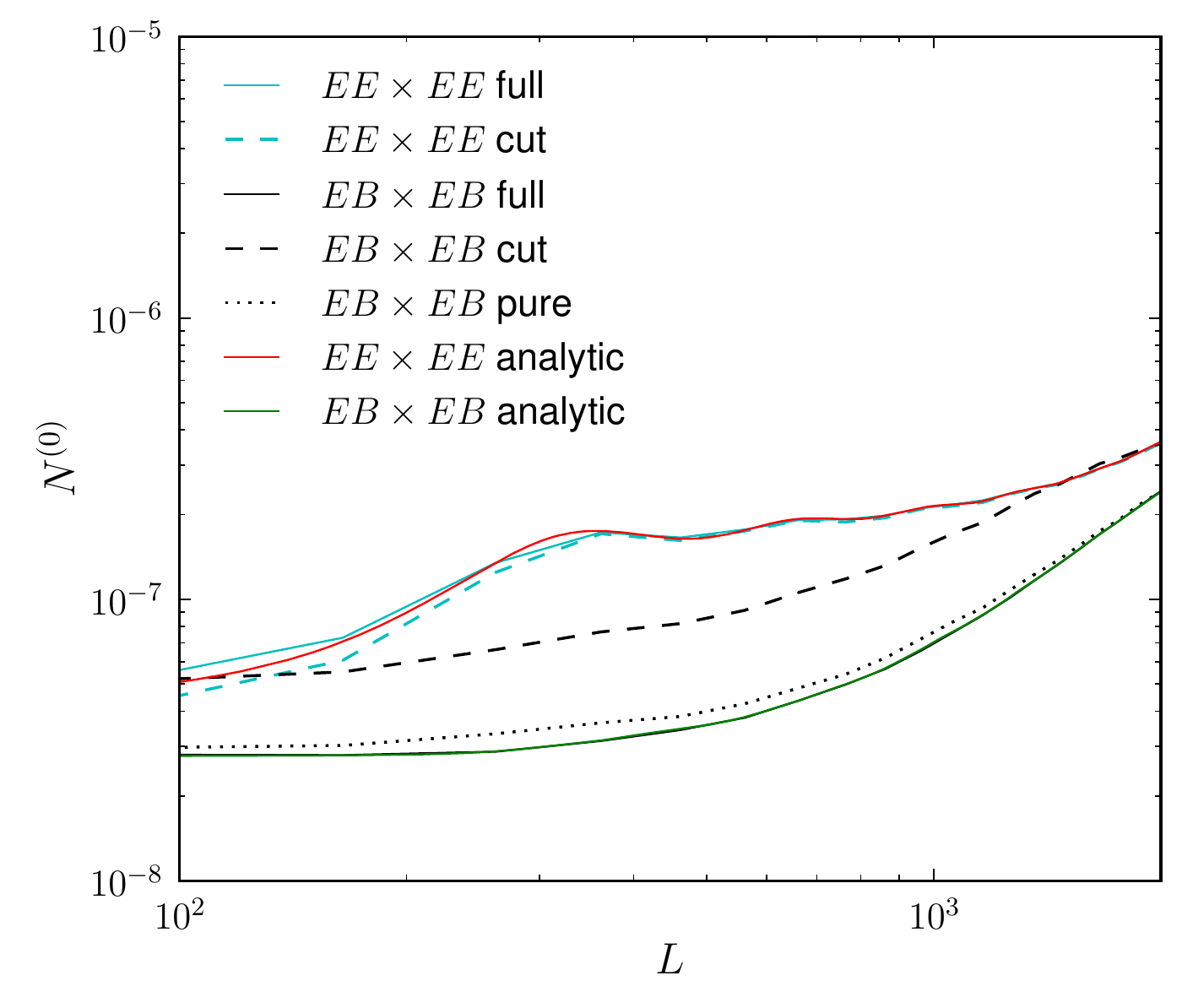}
\includegraphics[width=3.5in]{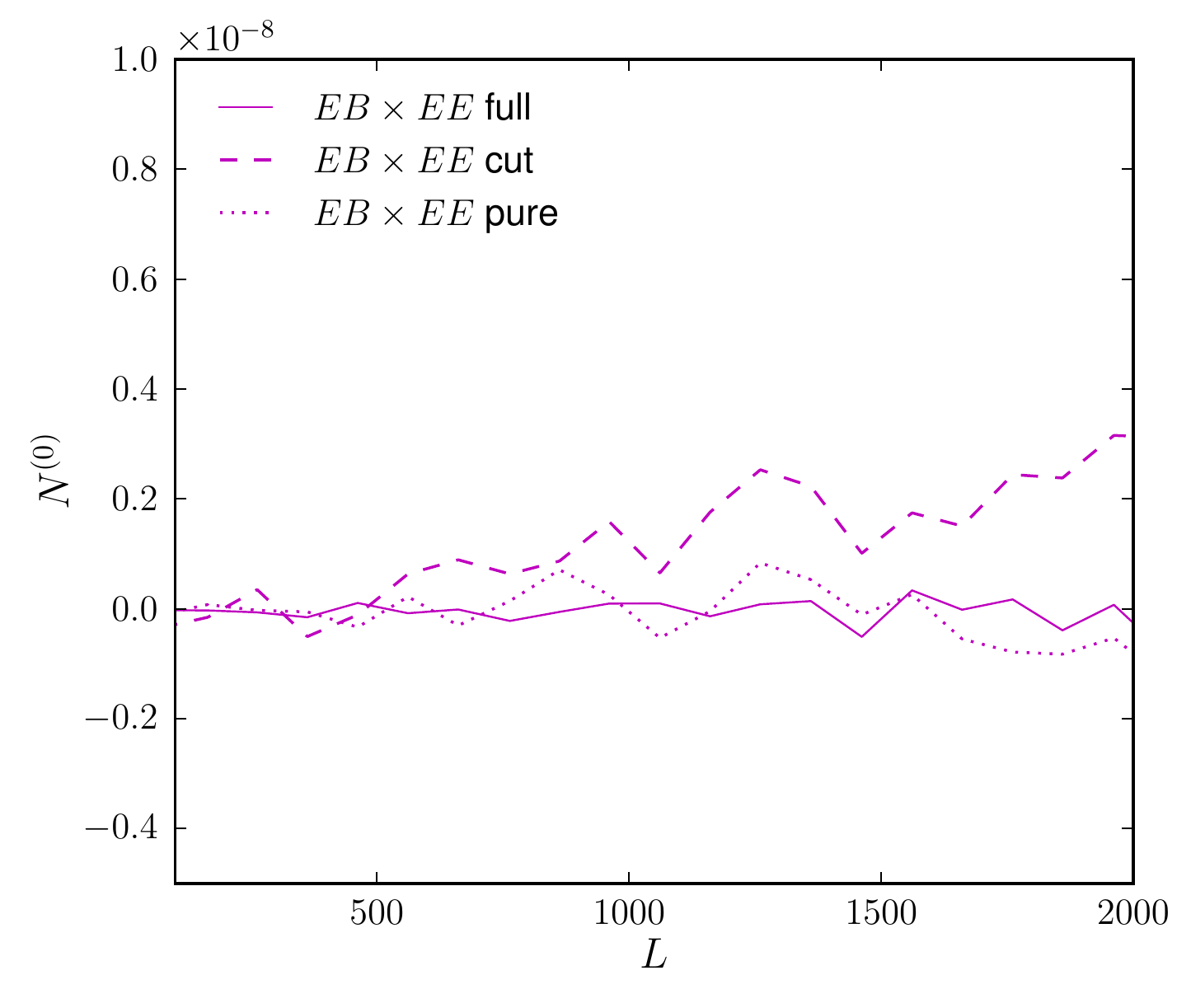}
}
\caption{
A comparison of the $\NZero$ bias reconstructions on the periodic and non-periodic sky patches.  Note that non-periodic cut sky power spectra have been scaled by an $f_{W4}$ factor.  Left:  The $EE\times EE$ and  $EB\times EB$ reconstructions.  Right:  The $EB\times EE$ reconstructions.
}
\label{fig:N0s}
\end{figure*}

Fig.~\ref{fig:N0s} shows a comparison of the $\NZero_{\text{sim}}$ bias power for the $EE\times EE$, $EB\times EB$ and $EB\times EE$ reconstructions in the periodic boundary and cut sky cases.  The analytic $EB\times EE$  has $\NZero_{EBEE}=0$ in the ideal full-sky case, but this becomes non-zero when there is leakage from $E$ into $B$. Using the pure-$B$ modes successfully reduces $\NZero_{EEEB}$ back to a low level. For  the $EB\times EB$ reconstruction, using the pure-$B$ modes results in an $\NZero_{\text{sim}}$ bias roughly the same amplitude as in the ideal full-sky case (i.e. with periodic boundary conditions).  The cut-sky analysis without the use of pure-$B$ modes however produces a much larger $\NZero$ bias, because the leakage of $E$ into $B$ increases the reconstruction variance (which is sensitive to the larger pseudo-$B$-mode power spectrum).  For the $EE\times EE$ reconstruction, the $\NZero_{\text{sim}}$ bias on the cut sky appears to be slightly lower on large scales than for the periodic sky patch.  This is probably due to the approximate $f_{W4}$ scaling that we have used being inadequate on scales approaching the scale of  variation of the window (see further discussion below).  The $\NZero_{\text{sim}}$ bias on the periodic-sky $EE\times EE$ reconstruction appears slightly high on large scales, but this is only due to binning.

In Fig.~\ref{fig:errorbar} we show a comparison of the 1$\sigma$ error bars of the various reconstructions considered in this paper.  The smallest error bars come from the periodic sky reconstructions which use $B$ modes: the $EB\times EB$ and $EB\times EE$ reconstructions.  The $EB\times EE$ estimator does slightly better on small scales, since there is no $\NZero$ noise term dominating at small scales in this case.  The error in the cut-sky $EB\times EB$ case is much worse, as expected due to $E$/$B$ mixing.  However, using the pure-$B$ mode reconstruction improves the cut-sky error bars dramatically.  Note that the periodic sky error bars have been scaled by an area factor of two to have the same error as expected from a $4.5^{\circ}\times4.5^{\circ}$ periodic patch.  For the comparison shown, the pure-$B$ mode error bars are roughly the same as the periodic sky error bars, showing that the pure-$B$ method works very well to mitigate the loss from $E$/$B$ mixing.  In Fig.~\ref{fig:errorbar} the binning used is $\Delta L=200$ to reduce the correlation between bins (correlations could otherwise cause the $\fsky$-scaled cut-sky diagonal error bars to look better than the periodic case).  Equation~\eqref{eq:varcut} shows that on small scales, without any bin correlations or $E$/$B$ mixing effects, an increase in error bar size of $\sim$15\% is expected for the windowed sky patch compared to the un-windowed periodic patch.

\begin{figure*}
\centerline{
\includegraphics[width=3.5in]{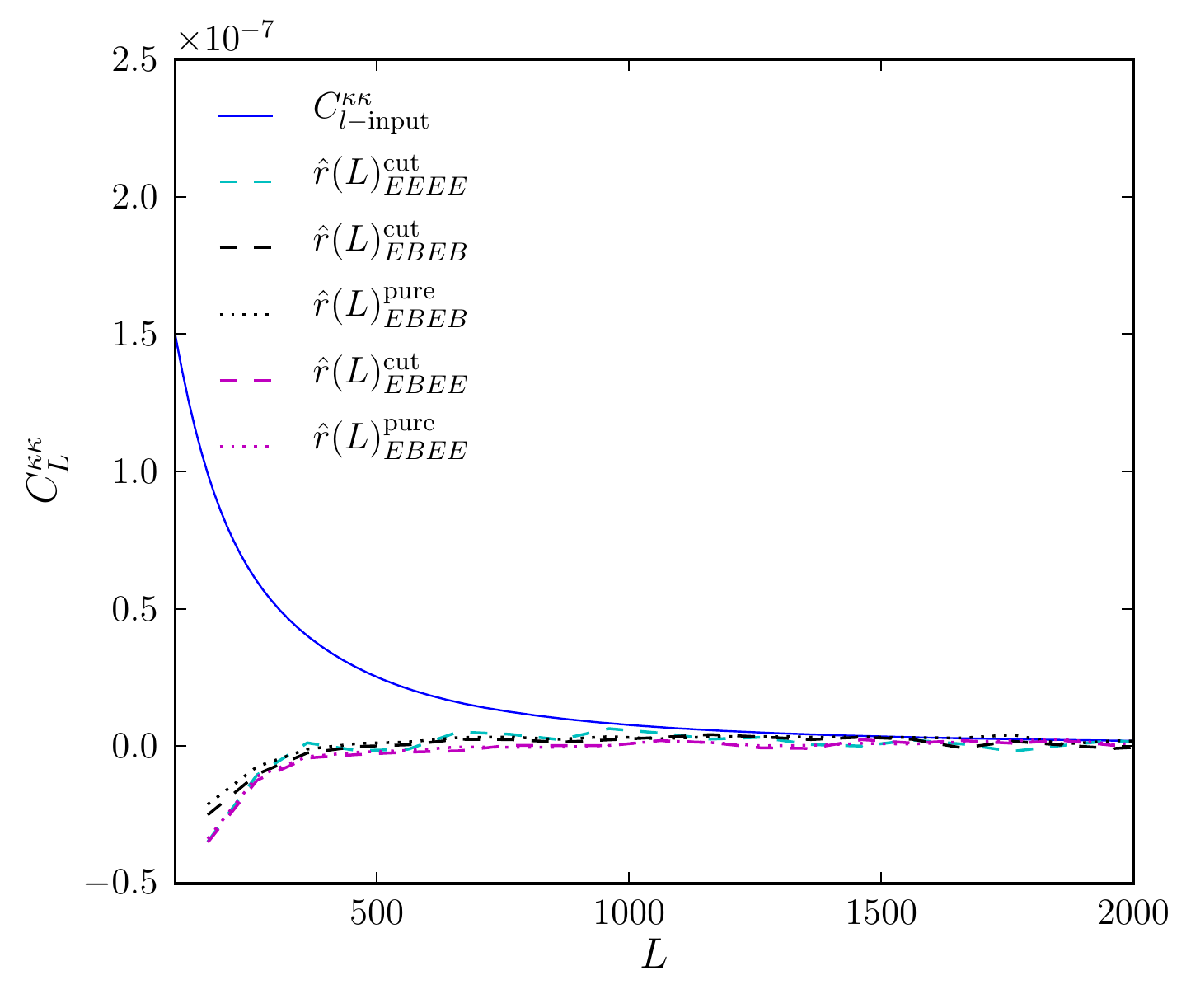}
\includegraphics[width=3.5in]{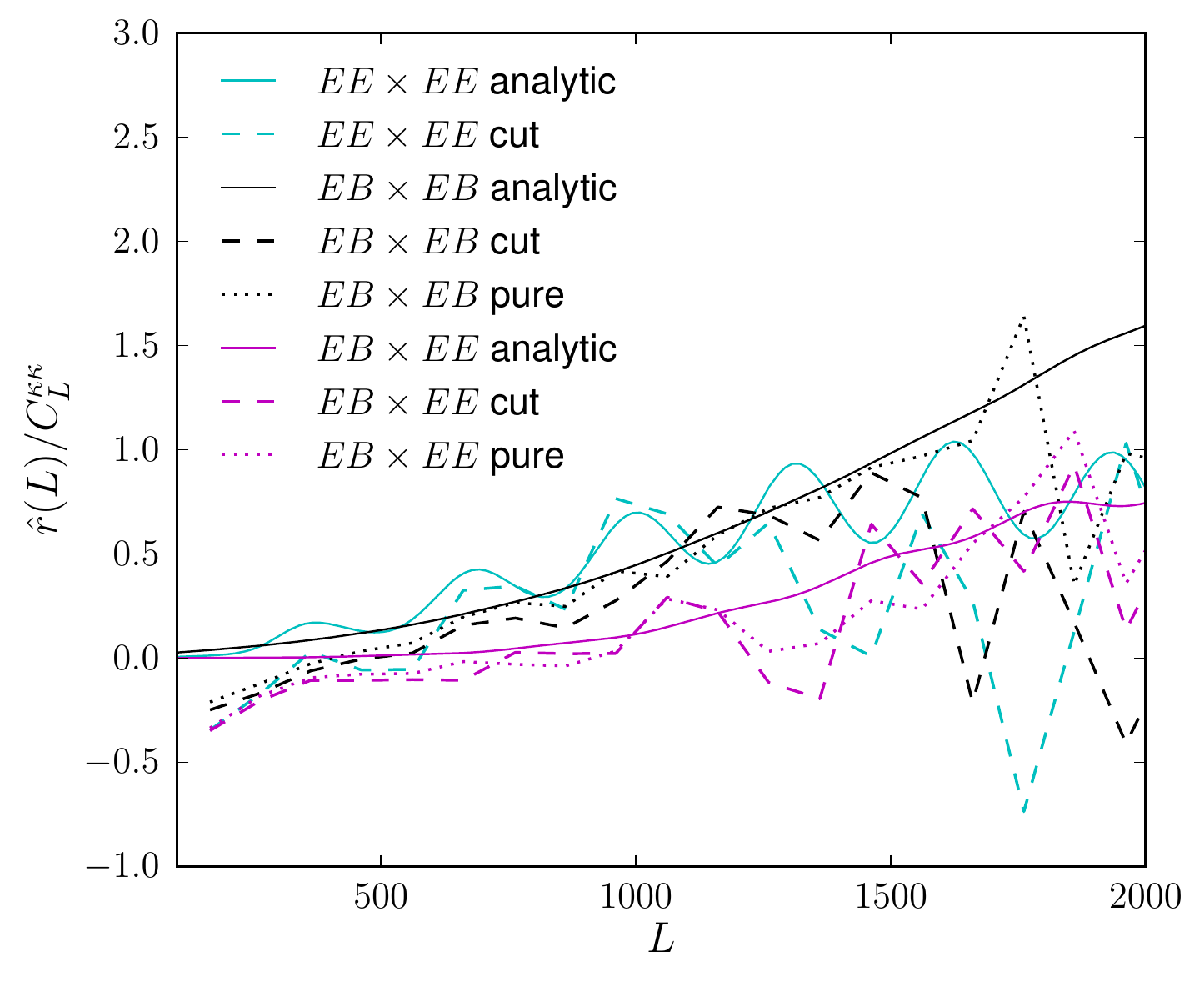}
}
\caption{
Left:  The residual biases of the cut sky and pure-$B$ mode sky reconstructions shown in Figs.~\ref{fig:cut} and \ref{fig:pure} compared to the input lensing power spectrum.  Right:  The ratio of the residual biases shown in the left-hand panel to the input lensing power spectrum. There is a significant negative bias on large scales, but the small-scale bias is fit quite well by the analytic $\NOne$ (Monte Carlo simulation error dominates on very small scales).
}
\label{fig:residual_biases_cut}
\end{figure*}

Finally, we investigate whether there are any additional biases that appear when using apodized cut sky maps and a simple constant $f_{W4}$ scaling factor. As shown in the left panel of Fig.~\ref{fig:residual_biases_cut}, we observe a substantial low bias on very large scales. Since the estimators by construction average to zero for Gaussian fields, any bias must scale at least with the lensing power $\Cdinput_l$, and we find a nearly-linear scaling consistent with Ref.~\cite{Ade:2013gez}.  The large-scale bias also affects the error bars of Fig.~\ref{fig:errorbar}, where on the largest scales the cut-sky error bars appear to be lower than in the optimal case without boundaries. This is because we have not recalibrated the errors for the bias.

To assess the size of the small-scale residual bias we show the ratio to the input lensing power spectrum in the right-hand panel of Fig.~\ref{fig:residual_biases_cut}.  On intermediate scales the analytic $\NOne$ is a reasonable fit to the bias. For our chosen noise level it is difficult to see the residual bias within the reconstruction noise on very small scales, so we also analysed a simulation with zero noise.  We found that the cut-sky residual bias was similar to the full-sky bias, and mostly still close to the analytic $\NOne$, with corrections small compared to the size of the signal.
This suggests that the residual bias on small scales is still fit reasonably well by the approximate analytic $\NOne$, although there may be some approximation error and also a mixing and/or scaling of reconstruction modes that is not accurately accounted for by the simple constant $f_{W4}$ diagonal scaling. For current-generation data the biases are small compared to the reconstruction noise, and the analytic $\NOne$ model is adequate except on large scales.

The large- and small-scale features of the residual bias are likely window dependent, and may be somewhat mitigated with a more optimal choice of window.
They can also be approximately modelled with an $L$-dependent normalization (transfer function).
However, accurate parameter estimation with more sensitive lensing reconstructions should consider a more detailed analysis of the full scale dependence of the window function effects relating the estimated and true lensing power spectra, including $L$-mixing due to the $\NOne$ bias as well as cut-sky effects.
More optimal estimators using full inverse variance weighting may turn out to have simpler properties than the simple windowed estimators considered here (though the $\NOne$ bias is non-local in $L$ and would still have to be modelled).

\section{Summary}

In this paper we simulate polarization lensing reconstruction for small areas of sky. We use these reconstruction simulations to investigate biases and signal-to-noise in both periodic and non-periodic windowed maps, and test the use of pure-$B$ modes in the standard quadratic estimator to mitigate the effects of $E$/$B$ leakage on the cut sky. The main findings are:
\\
\newpage
\textbf{For a periodic patch:}
\begin{itemize}
\item Analytic results for the $\NZero$ and $\NOne$ bias are adequate to model the leading reconstruction biases for current data. There is some evidence for small systematic deviations from the analytic results, possibly arising from higher-order effects or  assumed approximations, which may require more detailed study in future.
 \end{itemize}

\textbf{For the cut sky:}
\begin{itemize}
\item The large $\hd_{EB}$ mean field introduced by $E$/$B$ mixing is greatly reduced by using pure-$B$ modes in the estimator ($\hd_{EB}^{\rm pure}$), consistent with the low $EB$ mean field found by Ref.~\cite{Namikawa:2013xka}.
\item Using pure-$B$ modes significantly reduces the variance in the power spectrum reconstruction, and for the simple constant noise and nearly-constant window considered here, the reconstruction error is close to optimal.

\item We confirm the finding of~\cite{Ade:2013gez} that there is a substantial residual bias on large scales if a simple constant normalization is assumed.
\item The approach we present for reconstructing the lensing power spectrum on the cut sky
should be sufficient for current-generation CMB polarization measurements if the residual bias is accounted for by simulation, and makes a simple alternative to more numerically-costly, perturbatively-optimal estimators.
\item Detailed characterization of the normalization biases on the cut-sky may be required to fully exploit future more sensitive observations, where there may also be larger gains from the use of more optimal estimators (including going beyond perturbative leading order).

\end{itemize}

\label{summary}

\begin{acknowledgements}
RP thanks Prof.~Kuo at Stanford for initiating this work and hosting her while most of it was carried out.  RP acknowledges Wei-Hsiang Teng for providing an initial code framework.  RP acknowledges support from the Science and Technology Facilities Council via a research studentship. BDS thanks Chang Feng, Oliver Zahn and Alex van Engelen for discussions, and acknowledges support from a Miller Research Fellowship at Berkeley and a Charlotte Elizabeth Procter Honorific Fellowship at Princeton.
AL acknowledges support from the Science and Technology Facilities Council [grant number ST/I000976/1].

\end{acknowledgements}


\providecommand{\aj}{Astron. J. }\providecommand{\apj}{Astrophys. J.
  }\providecommand{\apjl}{Astrophys. J.
  }\providecommand{\mnras}{MNRAS}\providecommand{\aap}{Astron.
  Astrophys.}\providecommand{\aj}{Astron. J. }\providecommand{\apj}{Astrophys.
  J. }\providecommand{\apjl}{Astrophys. J.
  }\providecommand{\mnras}{MNRAS}\providecommand{\aap}{Astron. Astrophys.}

\end{document}